\documentclass[twocolumn]{aastex631}

\newcommand{\numax}{$\nu_{\rm max}$}
\newcommand{\msun}{$M_{\odot}$}

\shorttitle{Red Giant Masses}
\shortauthors{Pope et.~al.}
\graphicspath{{./}{figures/}}

\begin{document}

\title{TESS Asteroseismic Masses and Radii of Red Giants with (and without) Planets}

\author[0000-0001-7361-0828]{Myles Pope}
\affiliation{Howard University \\ 2400 6th St NW \\ Washington, DC 20059}
\affiliation{Johns Hopkins University \\ 3400 North Charles Street \\ Baltimore, MD 21218}

\author[0000-0001-5926-4471]{Joleen K. Carlberg}
\affiliation{Space Telescope Science Institute \\
3700 San Martin Dr.\\Baltimore, MD 21218}

\author[0000-0003-3305-6281]{Jeff Valenti}
\affiliation{Space Telescope Science Institute \\
3700 San Martin Dr.\\Baltimore, MD 21218}

\author[0009-0009-7822-7110]{Doug Branton}
\affiliation{University of Washington, 1400 NE Campus Parkway, Seattle, WA, 98195-4550}

\correspondingauthor{Joleen Carlberg}
\email{jcarlberg@stsci.edu}



\begin{abstract}

We present a study of asteroseismically derived surface gravities, masses, and radii of a sample of red giant stars both with and without confirmed planetary companions
using TESS photometric light curves. These red giants were drawn from radial velocity surveys, and their reported properties in the literature rely on more traditional methods using spectroscopy and isochrone fitting. 
Our asteroseismically derived surface gravities achieved a precision of $\sim$0.01 dex; however,  they were on average $\sim$0.1~dex smaller than the literature.
The systematic larger gravities of the literature could plausibly present as a systematic overestimation of stellar masses, which would in turn lead to overestimated planetary masses of the companions. 
We find that the fractional discrepancies between our asteroseismically-determined parameters and those previously found are typically larger for stellar radii ($\sim$10\% discrepancy) than for stellar masses ($<5$\% discrepancy). However, no evidence of a systematic difference between methods was found for either fundamental parameter.
Two stars,  HD~100065 and HD~18742, showed significant disagreement with the literature in both mass and radii.  We explore the impacts on updated stellar properties on inferred planetary properties and caution that red giant radii may be more poorly constrained than uncertainties suggest.

\end{abstract}



\section{Introduction} \label{sec:intro}
A star's mass is one of its most fundamental properties. It is the dominant driver of its internal structure and evolution, and it   influences the formation and dynamics of any exoplanets orbiting the star. 
Before the advent of spaced based, high precision photometric monitoring of larger samples of stars, including red giants, by the \textit{Kepler} \citep{Borucki2010} and TESS \citep{TESS2015} missions, masses of isolated red giant stars were inferred through a combination of spectroscopy and broadband photometry, each with its limitations.  Broadband photometry allows for the fitting of the full spectral energy distribution of a star, yielding a stellar temperature, radius, and luminosity to compare to stellar isochrones or evolutionary tracks. The accuracy in these fits are influenced by the accuracies of the stellar distance (a limiting factor in the pre-Gaia era), bolometric corrections, stellar reddening estimate, and metallicity indicator.
Spectroscopy allows for a distance- and reddening-agnostic determination of the stellar temperature, $\log g$, and metallicity  by solving for the stellar atmospheric properties that reproduce the equivalent widths of a wide range of stellar absorption features, typically \ion{Fe}{1} and \ion{Fe}{2} lines, often (and most easily) under the assumption of local thermodynamic equilibrium (LTE). The accuracy and precision of these methods are limited by uncertainties in the equivalent width measurements (from, e.g., line blending and continuum placement errors), simplifying assumptions in the 1D LTE modeling, and inaccurate or incomplete atomic line information.
With a stellar temperature, $\log g$ and radius, the stellar mass can be solved for directly. Alternatively, a suite of photometric and/or spectroscopic constraints  can be compared to model stellar evolution tracks to yield estimated masses, using Bayesian inference, such as PARAM \citep{2006A&A...458..609D}, \texttt{isoclassify} \citep{2017ApJ...844..102H,2020AJ....159..280B}, and \texttt{kiahoku} \citep{2020ApJ...888...43C}. 
These mass estimates will differ depending on the selection of underlying suite of models and their associated choices of physical prescriptions.
Even when the physical prescriptions are common between codes, \cite{2020A&A...635A.164S} demonstrated that the temperature and luminosity predicted for red giants at a given mass can differ by as much as 40~K in temperature and 10\% in luminosity. Furthermore, the mass uncertainty in some regions of Hertzsprung-Russell and Kiel diagrams can become multi-modal, as stars of different masses pass through the same narrow band of observed parameter space multiple times \citep[e.g., the luminosity bump,][]{2015MNRAS.453..666C}.

In contrast, asteroseismology allows a much more precise determination of $\log g$ and density (and thus mass and radius) via scaling relations \citep{1995A&A...293...87K} if high-precision photometric monitoring of sufficient baseline and time sampling are available. Stars along the red giant branch (RGB) typically have $\log g\sim$1--3.5, which corresponds to typical peak oscillation frequencies of $\sim1.4$--$380~\mu$Hz, or oscillation periods of  $\sim 8$ days (upper RGB) down to $\sim$45 min (lower RGB). 
The highly successful \textit{Kepler} mission led to the systematic measurement of asteroseismic properties for tens of thousands of red giant stars \citep{2010A&A...522A...1K,2013ApJ...765L..41S,2018ApJS..236...42Y} revolutionizing our understanding of these stars, particularly when leveraging partnerships with large spectroscopic surveys to provide precise and homogeneously derived stellar temperatures.  These partnerships also paved the way for better understanding the limitations in both our spectroscopic and asteroseismic methods. For example, the APOKASC partnership  leveraged \textit{Kepler} asteroseismology to apply systematic corrections to the spectroscopic $\log g$ measurements of the APOGEE surveys \citep{meszaros_calibrations_2013,2014ApJS..215...19P}. In return, incorporating APOGEE measured abundances and temperatures of red giants tied to open clusters allowed an exploration of differences between asteroseismic pipeline results \citep{pinsonneault_second_2018}.

While \textit{Kepler} revolutionized the study of red giant populations, its limited sky coverage meant that its benefits could not be applied to specific populations of interest if they fell outside \textit{Kepler}'s coverage. One notable population consists of the brightest, best studied red giants, which are spread across the sky.
The advantage of the TESS mission over \textit{Kepler} is broad  sky coverage, where, during the nominal 2 year mission, a sector of sky is observed every 30 min (the full frame images) for  an average of 27 days. 
Red giant oscillation frequencies are well within this typical TESS observing cadence, and many bright giants are additionally observed at the higher 2 minute observing cadence and may fall in more than one sector, increasing the total baseline of observation. Furthermore, beginning with sector 28 in year 2, full frame images were sampled every 10 min, and select targets were sampled at either 2 minute or 20 secs\footnote{\url{https://tess.mit.edu/public/target_lists/target_lists.html}}.  Large asteroseismic studies of red giant oscillations capitalizing on this rich dataset have already begun  \citep[e.g.,][]{hon_quick_2021,2022AJ....164..135H} and are expected to continue. 

Asteroseismology has already played an important role in constraining the properties of evolved planet hosts, notably in the controversy of the masses of the ``retired A stars'' \citep{2007ApJ...665..785J,2011ApJS..197...26J}. These cool, moderately evolved stars occupy a region of the Hertzsprung-Russell diagram where mass discrimination should be easier than further up the red giant branch. 
These stars' intermediate masses ($M \gtrsim 1.3$~\msun) have been challenged on the basis that both their rotational velocities \citep{2011ApJ...739L..49L,2013ApJ...774L...2L} and space velocity dispersion \citep{2013ApJ...772..143S} are more akin to 1-1.2~\msun\ stars. These claims have been refuted by \cite{2013ApJ...763...53J} and \cite{2018ApJ...860..109G}, the latter of whom argued that at most the masses may be overestimated by $0.12$\msun, maintaining their classification as ``retired A'' stars.
\cite{2017MNRAS.469.1360C} used \textit{Kepler} to revise the masses of two stars, one up and one down. \cite{2017MNRAS.472.4110S} asteroseismically studied eight of these evolved stars with the SONGS telescope, and found masses to be overestimated by 15-20\%. More recently \cite{2021AJ....162..211H} derived an  asteroseismic mass of $\iota$~Dra, of $1.54\pm$~0.09~\msun, which is a substantial revision up from the \cite{2013A&A...557A..70M} value of $1.14\pm0.16$~\msun\ and lower than their own ``empirical'' (spectroscopic $\log g$ and photometric radius) measurement of $1.72\pm0.29$~\msun.

In this work, we leverage this broad sky coverage to measure $\log g$, mass, and radius for a sample of red giant stars that have been spectroscopically monitored for planetary companions. We have already homogeneously  measured spectroscopic $\log g$ for this specific sample of red giants and have collected independent $\log g$, mass, and radii measurements from different literature source to allow a cross-comparison of stellar properties derived with different methodologies. We obtain TESS light curves for our sample and measure asteroseismic parameters for stars whose oscillation frequency spectra are of sufficient quality. With these results, we compare $\log g$, radius, and mass inferred from asteroseismology to other methodologies to investigate the accuracy of the previously determined red giant star and planetary systems.

\section{Sample Selection}  \label{sec:sample}
The stars in this sample primarily reside in the Southern Hemisphere and were selected from radial velocity surveys of red giants \citep{johnson_eccentric_2006,PTPS,2010ApJ...725..875I,frink01}.  Our sample includes stars both that have had planets found around them  and those that have no known companions,  and the sample is further limited to a subset that have been followed up for a detailed spectroscopic abundance analysis (\citealt{2020AAS...23527403B}, Carlberg et.~al. in prep) and thus have a homogeneous 
spectroscopic determination of $\log g$, $T_{\rm eff}$, and [Fe/H].
That full sample of stars consists of 39 planet hosts and 47 non-hosts.  However, of these only 22 of the planet hosts (see Table \ref{tab:sample}) and 26  of the non-hosts (see Table \ref{tab:sample2}) have TESS data amenable to asteroseismic analysis.
For the planet hosting stars, we additionally retrieved 
 $\log g$, mass and radii from Exo.MAST\footnote{\url{https://Exo.MAST.stsci.edu}}, which, by default, supplies metadata from the NASA Exoplanet Archive \citep{2013PASP..125..989A}, an actively maintained database of exoplanetary systems and properties published in the literature. Exo.MAST can also supply meta data from the Exoplanetary Orbit Database \citep{2014PASP..126..827H}, which was actively maintained through June 2018. Because the data in Exo.MAST comes from many different published studies, this comparison dataset is very heterogeneous in nature but represents common techniques for deriving stellar parameters. For a more homogeneous comparison, we also compare  $\log g$, mass and radii for the planet hosts found in the SWEET-Cat 2.0 catalog \citep{2021A&A...656A..53S}, which is a catalog aimed at deriving homogeneous stellar parameters from high quality spectra combined with Gaia eDR3 parallaxes.

\begin{deluxetable}{cccc}[t!]
\tablenum{1}
\tablecaption{Red Giant Planet Host Sample \label{tab:sample}}
\tablewidth{0pt}
\tablehead{
\colhead{Star Name} & \colhead{Sectors} & \colhead{Exposure Time} & \colhead{Pipeline\tablenotemark{a}} \\
\colhead{} & \colhead{} & \colhead{(s)} & \colhead{} 
}
\startdata
18 Del\tablenotemark{b} & 55 & 120 & 1\\
24 Sex & 8,35,45,46 & 120 & 1\\
  & 8 & 1800 & 2,3\\
  & 35 & 600 & 2,3\\
7 CMa & 6,33 & 120 & 1 \\
& 6 & 1800 & 2,3 \\
& 33 & 600 & 2,3\\
$\alpha$ Ari & 17,42,43 & 120 & 1\\ 
\enddata
\tablenotetext{a}{1- SPOC, 2- TESS-SPOC, 3-QLP, 4-TASOC, 5-CDIPS}
\tablenotetext{b}{Data became available after our original cutoff date but was included to allow for an expanded literature comparison in Section \ref{sec:validate}}
\tablecomments{Table 1 is published in its entirety in the machine-readable format.
      A portion is shown here for guidance regarding its form and content.}
\end{deluxetable}

\begin{deluxetable}{cccc}[t!]
\tablenum{2}
\tablecaption{Red Giant Control Sample\label{tab:sample2}}
\tablewidth{0pt}
\tablehead{
\colhead{Star Name} & \colhead{Sectors} & \colhead{Exposure Time} & \colhead{Pipeline\tablenotemark{a}} \\
\colhead{} & \colhead{} & \colhead{(s)} & \colhead{}
}
\
\startdata
HD105096 & 22,49 & 120 & 1\\
& 22 & 1800 & 2,3 \\
HD108991 & 22,49 & 120 & 1\\
& 22 & 1800 & 2,3 \\
HD115202 & 10 & 120 & 1\\
& 10 & 1800 & 2,3 \\
HD121056 & 11,38 & 120 & 1\\
& 11 & 1800 & 2,3 \\
& 38 & 600 & 2,3\\
\enddata
\tablenotetext{a}{1- SPOC, 2- TESS-SPOC, 3-QLP, 4-TASOC, 5-CDIPS}
\tablecomments{Table 2 is published in its entirety in the machine-readable format.
      A portion is shown here for guidance regarding its form and content.}
\end{deluxetable}

\section{TESS Data Analysis}\label{sec:data}
\subsection{TESS Data Acquisition}\label{sec:get_tess}
Our astersoseismic measurements rely on the \texttt{Lightkurve} package \citep{2018ascl.soft12013L} to both retrieve and analyze TESS light curves.
 TESS photometric light curves were gathered from the MAST archive using the search function within the \texttt{Lightkurve} package. For this analysis, the lightcurves that were used were a mix of short cadence  and long cadence observations.
 Prior to $\sim$ mid-2020, short cadence and long cadence corresponded to 120~s and 1800~s, respectively. Afterward,  the long cadence  was reduced to 600~s and short cadences of both 20~s and 120~s were available for select targets.
 Later changes to the cadence were made beginning with Sector 56, but these later data were not yet available during our analysis. 
 The available lightcurves were processed by one or more pipelines, with corresponding High Level Science Product lightcurves available from MAST directly from the \texttt{Lighktkurve} package. Each pipeline used a different aperture mask on the original target pixel file from a given sector, leading to a slightly different lightcurve. The majority of stars were processed by the Quick Look Pipeline  \citep[QLP, ][]{QLP1,QLP2,QLP3,QLP4,QLP_HLSP}.  However, we found the lightcurves  produced by this pipeline were generally too noisy for precise asteroseismic measurement.  Our preferred lightcurves were those produced by the  SPOC \citep{SPOC}  and TESS-SPOC \citep{TESS-SPOC,TSPOC_HLSP} pipelines. Two additional  pipelines TASOC \citep[][]{TASOC1,TASOC2,TASOC_HLSP} and CDIPS \citep{CDIPS,CDIPS_HLSP} were available for use in some cases. However, at the time of our analysis, fewer sectors had been processed by these pipelines compared to SPOC and TESS-SPOC, leading to noisier light curves.
 In Tables  
 \ref{tab:sample} and \ref{tab:sample2}, we list for each star which TESS sectors have available lightcurves from at least one the pipelines above. Each row for a given star lists the sectors for which homogeneous data (same exposure time and pipeline) is available.

If a target had observations in multiple sectors that were produced by the same pipeline with the same observation cadence, the lightcurves were stitched together in order to get a longer baseline for analysis. This proved to be vital, as the TESS 27.4 day observation period only covers a few oscillation cycles of the most evolved stars.
In general, the short cadence data was preferred to the long cadence data, as the finer time sampling  allowed for better resolution of  peaks in the autocorrelation functions, which in turn led to more precise masses.   For the measurements of $\Delta \nu$ and \numax\ described in the next sections, we processed the available data from each combination of pipeline and cadence independently. We discarded datasets with clearly inferior quality (as noted, usually the QLP and TASOC pipelines). When multiple independent analyses gave results of comparable quality, the final results were averaged.

\subsection{Measuring $\nu_{\rm max}$}
\begin{figure}
\centering
\includegraphics[width=0.45\textwidth,trim={1cm 0 1cm 0}]{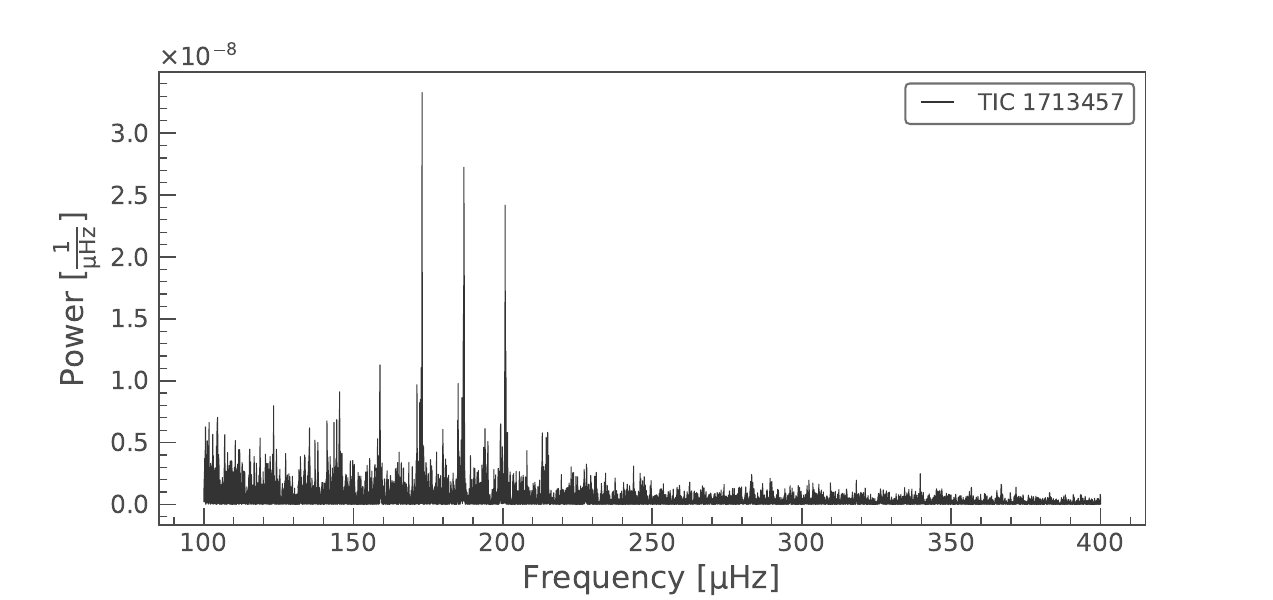}
\caption{Normalized frequency-power spectrum of 24 Sex 2 minute cadence data. \label{fig:power}}
\end{figure}

The measurement of \numax\ was done with the built-in methods of the \texttt{Lightkurve} package.
We started by accessing the lightcurve data as detailed in the previous section. The lightcurves that we obtained had multiple flux columns, and the ones we used for the analysis were the normalized  Pre-search Data Conditioning Simple Aperture Photometry (PDCSAP) flux columns and the TESS Barycentric Julian Day (BJTD) time. The lightcurve was transformed into a periodogram using the Lomb-Scargle method, the default of the \texttt{to\_periodogram} method in the \texttt{LightCurve} class. When generating the periodogram, we sometimes used an oversample factor depending on how much data we had. If we had multiple sectors of data, then the oversample would not be needed since there were enough independent points to make a full spectrum, but in the case of objects where we only had one sector, the oversample factor was needed to get a full spectrum. An example periodogram is shown in Figure \ref{fig:power}. We bounded the periodogram by a minimum and maximum frequency that was $\sim 100$~$\mu$Hz  above and below the expected $\nu_{\rm max}$ calculated with the  spectroscopic $\log g$ and temperature measurements (see Section \ref{sec:results} and Equation \ref{eq:1}). The periodogram gave us a spectrum of power versus frequency, which was then flattened by dividing the periodogram by a background estimate using a moving filter in $\log_{10}$ frequency space. 
The filter width reduced the power of the frequencies at the lower end of the spectrum in order to reduce noise, but also caused the stellar signal to be reduced at that frequency.
The default filter width of 0.01 (in units of $\log_{10}$ frequency [$\mu$Hz]) was suitable for stars with $\log g \geq 2.5$.  For stars that had a $\log g < 2.5$, we increased the filter width to minimize the impact to the stellar signal. For stars with $2 \leq \log g < 2.5$, we used a filter width of 0.75, and for $\log g < 2$, we used a filter width of 1. Stars that oscillated at a lower frequency still proved troublesome even with the larger filter width, because it was hard to distinguish the noise from the actual signal at low ranges. This mainly comes from the fact that the low frequency oscillators would only make complete oscillations a few times an observation, which would give the oscillation a power that would be as low as a noise.

\begin{figure}
\centering
\includegraphics[width=0.5\textwidth]{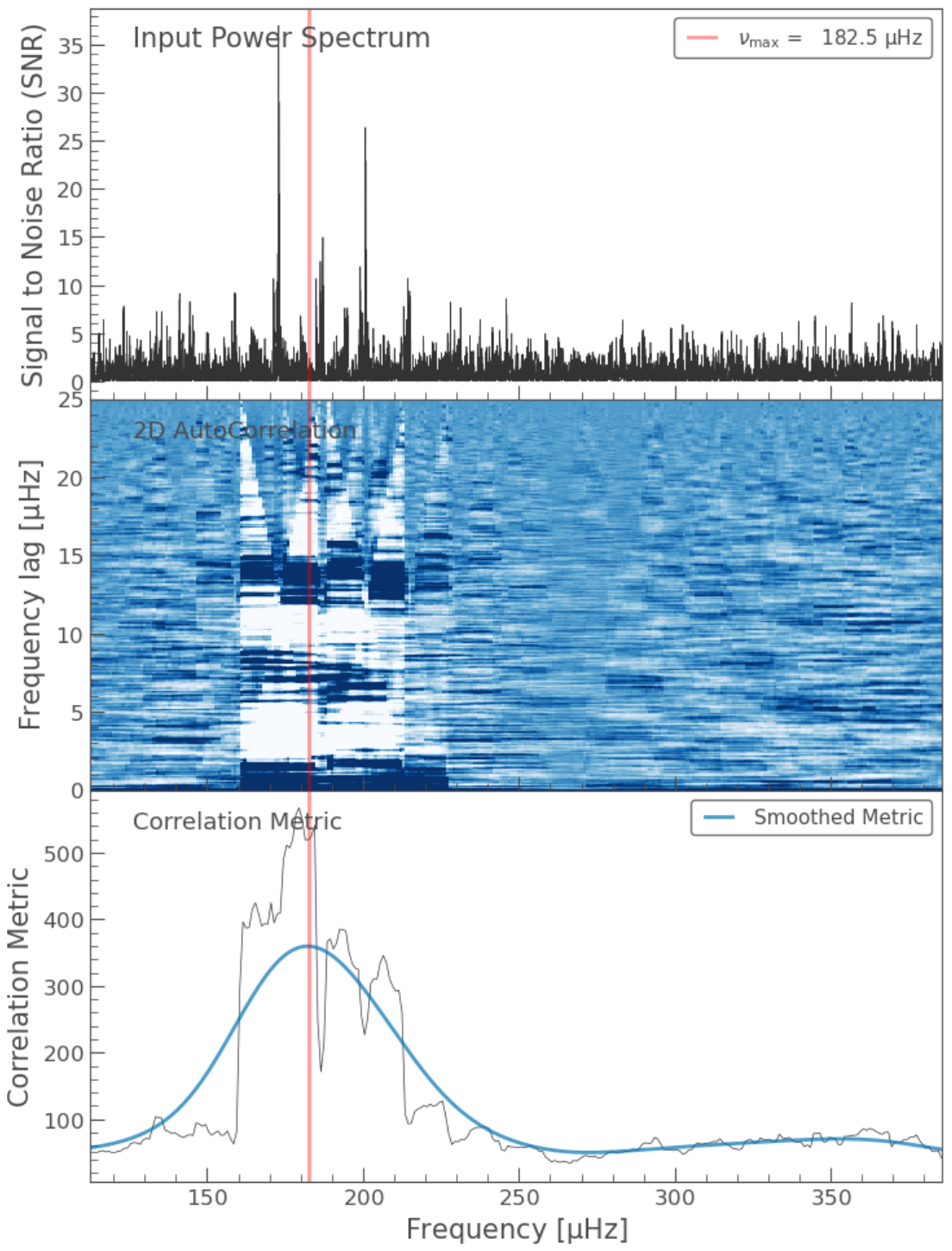}
\caption{Diagnostic plots from \texttt{Lightkurve}'s \texttt{diagnose\_numax()} function illustrating the derivation of \numax. (Top): The filtered, normalized periodogram. (Middle) The 2D auto-correlation function. (Bottom) The peak of the smoothed correlation metric yields \numax. \label{fig:numax}}
\end{figure}

After reducing the noise in the periodogram, we passed the periodogram to the \texttt{estimate\_numax} method, which computes a 2D auto-correlation and fits a Gaussian to the correlation metric. The frequency of this Gaussian's peak is \numax, and Figure \ref{fig:numax} shows an example of the diagnostic plots produced by \texttt{Lightkurve} to determine the quality of the fit.
 This is where the longer baseline was crucial, as targets with a single observation tended to lack a dominant signal that would indicate a clear $\nu_{\rm max}$.
In a few cases, the alias of the stellar frequency (at 2x the true frequency) resulted in comparable or larger power than the lower peak. However, these were relatively easy to identify since our spectroscopic $\log g$ could be used to discriminate the more likely set of true frequencies.  

\subsection{Calculating $\Delta\nu$ }
After the $\nu_{\rm max}$ was found, we measured the characteristic frequency separation, $\Delta\nu$. The $\Delta\nu$ was estimated using a 1-D auto-correlation function on the periodogram, with a window centered at the $\nu_{\rm max}$ and with a width that was half of the full width half maximum of the Gaussian fit in the \numax\ derivation step. The $\Delta\nu$ was then selected using scipy's \texttt{find\_peaks} function, within a window that was between 0.75 and 1.25 of the empirical $\Delta\nu$, which was calculated using the equation for red giants. 

\begin{figure}[b!] 
\centering
\includegraphics[width=0.5\textwidth, trim={0 5cm 0 5cm}]{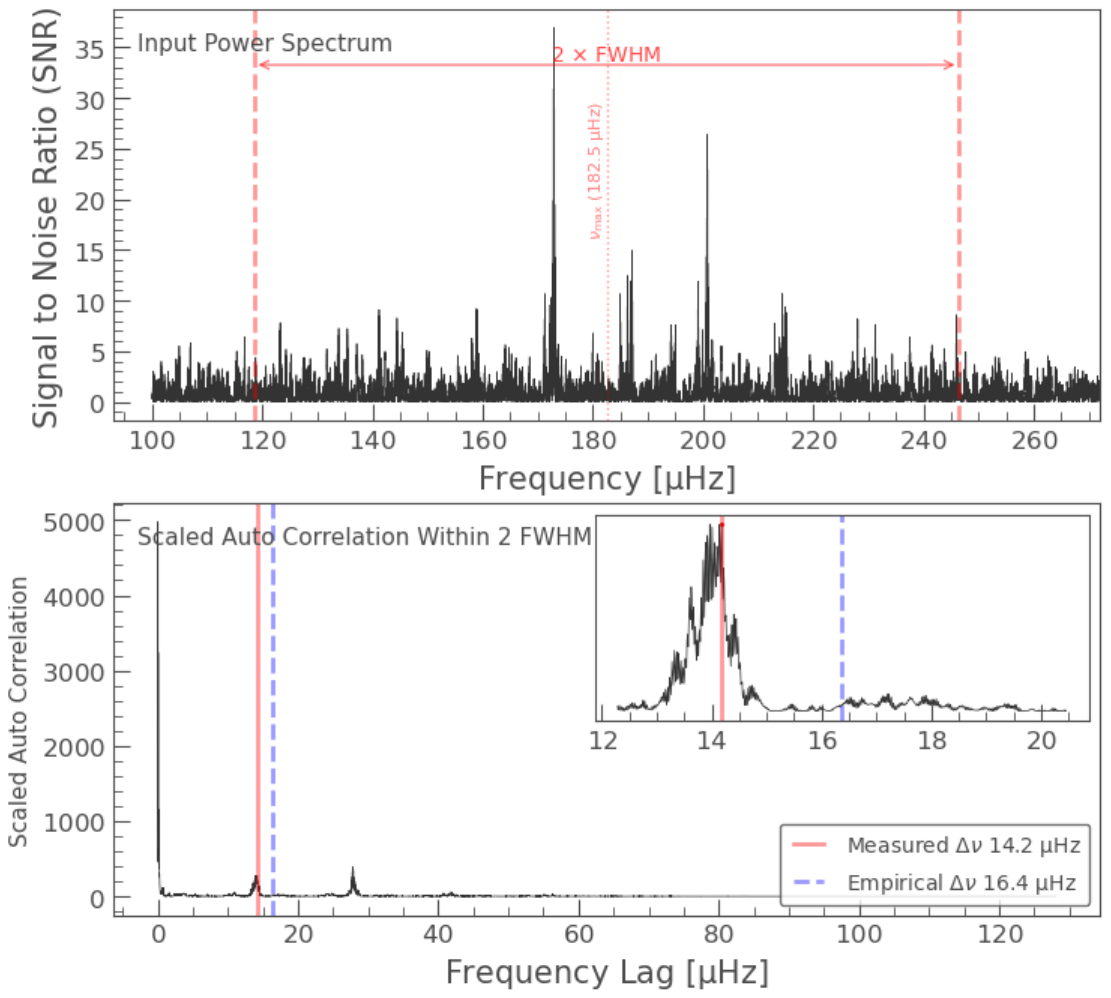}
\caption{Diagnostic plots from \texttt{Lightkurve}'s \texttt{diagnose\_numax()} function illustrating the measurement $\Delta\nu$. (Top): The auto-correlation is restricted to the frequency region within 2 full-width at half maximum (FWHM) of the Gaussian around \numax. (Bottom): $\Delta\nu$ is chosen as the largest peak in the auto-correlation function that is near the empirical $\Delta\nu$, estimated from the \numax. In this example, a better estimate of $\Delta\nu$ could be found by fitting the correlation function as illustrated in Figure \ref{fig:delnu2}\label{fig:deltanu}}
\end{figure}

\begin{figure*}
\includegraphics[width=1.0\textwidth,trim={0 4cm 0 4cm}]{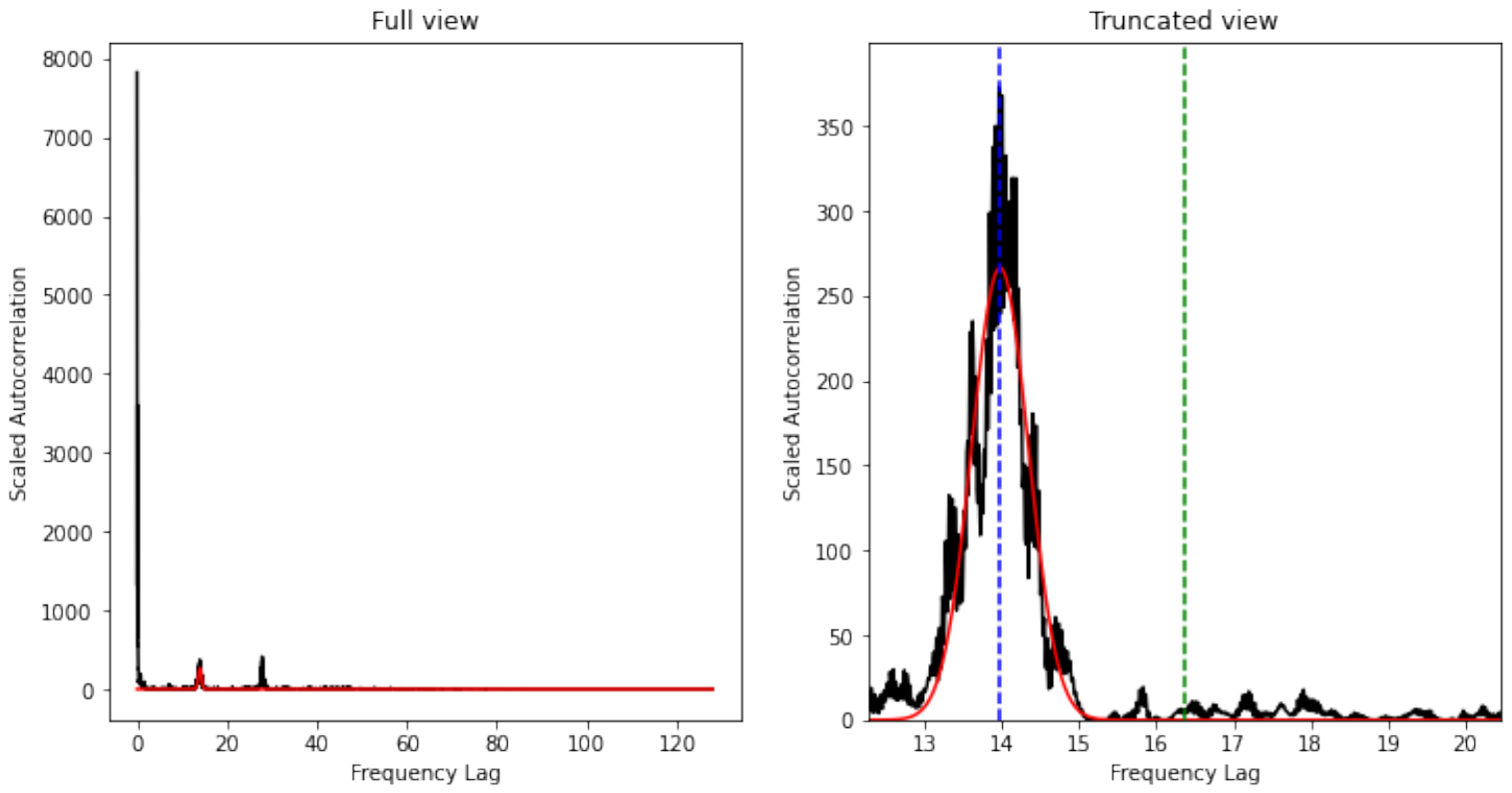}
\caption{Gaussian fit to the auto-correlation function of 24 Sex calculated by \texttt{Lightkurve}. This is the same data plotted in the bottom panel of Figure \ref{fig:deltanu}. We use the frequency lag at the peak of the fitted Gaussian function rather than the single highest point in the auto-correlation function to define the location of $\Delta \nu$, leading to a slightly smaller $\Delta \nu$ than what is returned by \texttt{Lightkurve}. \label{fig:delnu2}}
\end{figure*}

\texttt{Lightkurve} simply returns the frequency of the highest peak in the window as the measured $\Delta\nu$, with the expectation that peaks at lower correlation power are due to correlations between stellar oscillation frequencies of different spherical order (e.g., $l=1$ frequencies with $l=2$ frequencies).  However, we found (particularly with the long cadence data) that the auto-correlation peaks of the same spherical order and different spherical order were not always resolved. Additionally, noisy periodograms also led to noise autocorrelations with spurious peaks superimposed on the dominant peak of interest. Figure \ref{fig:deltanu}  shows an extreme example of both of these effects. The auto-correlation peak  of interest is centered just below 14 $\mu$Hz, but the noise spike at 14.02 $\mu$Hz is identified as the best fit.
 For this reason, we recalculated the auto-correlation function in the same way as \texttt{Lightkurve} but we additionally fit a Gaussian curve to better refine the $\Delta\nu$ frequency in the presence of noisy data. 
  The mean value of that Gaussian was taken as the $\Delta\nu$.  Figure \ref{fig:delnu2} shows the example of the refined fit to auto-correlation function of Figure \ref{fig:deltanu}, yielding a $\Delta\nu$ of 13.98 $\mu$Hz). In most cases the $\Delta\nu$ obtained using \texttt{Lightkurve}'s default method and the $\Delta\nu$ we found using the Gaussian had a very small difference, but the improvement allowed us to get a more precise mass and radius measurement.

As a final check  of the measured $\Delta\nu$, we inspected the  echelle  diagrams, which are vertically stacked slices of the periodogram  that are sliced in widths of length $\Delta\nu$.  We expect to see vertical structure in the echelle diagram, corresponding to the $l = 0, 1, $ and 2 radial modes of pulsation, whereas a $\Delta\nu$ corresponding to spurious noise peaks generally showed a single strong correlation peaks with no obvious structure. 
 We discarded any  $\Delta\nu$ that did not show evidence of vertical structure, which we checked by eye. We used this as both a wellness check and to discern between conflicting $\Delta\nu$ measurements.

\subsection{Uncertainties}
We quantify our typical uncertainties using the subset of stars for which we have more than one measurement by taking the standard deviation of measurements for individual stars averaged over the sample.  For \numax, this corresponds to 19 stars, and we find typical uncertainties of 0.67~$\mu$Hz for \numax$< 150$~$\mu$Hz and  2.47~$\mu$Hz for \numax$> 150$~$\mu$Hz.
For $\Delta\nu$, the typical uncertainty is  0.06~$\mu$Hz for all 17 stars with multiple measurements.   These uncertainties are formally propagated to mass, radius and $\log g$ using the \texttt{uncertainties} Python package, adopting a typical uncertainty in the spectroscopic temperature of $80~K$.

We do note that the largest potential source of error in the $\Delta\nu$ measurement is the potential mis-identification of the peak auto-correlation frequency. However, the use of echelle diagrams to discard potential spurious results should reduce this likelihood. The discarded $\Delta\nu$ measurements  were typically 4--8\%\ discrepant with the adopted ones. Additionally, we discarded $\Delta\nu$ if the measurement quality was questionable and it resulted in an unphysical low mass measurement ($\lesssim0.85\rm{M}_\odot$). Our validation with literature measurements in Section \ref{sec:validate} gives us confidence that our vetting procedure would have discarded low confidence measurements. 

\begin{rotatetable*}
\begin{deluxetable*}{lrrrrrr|rr|rrr}
\tablenum{3}
\tablecaption{Planet Host Sample Final Results\label{tab:results}}
\tablewidth{0pt}
\tablehead{
\colhead{} & \multicolumn{6}{c}{Asteroseismic} & \multicolumn{2}{c}{Spec} & \multicolumn{3}{c}{Exo.MAST} \\ 
\colhead{Star} & \colhead{Mass} & \colhead{Radius} & \colhead{$\log g$} & \colhead{Method} & \colhead{$\nu_{\rm max}$} & \colhead{$\Delta\nu$} &
\colhead{$T_{\rm eff}$} & \colhead{$\log g$} & 
\colhead{$\log g$} & \colhead{Mass} &\colhead{Radius} \\
\colhead{} & \colhead{($M_{\sun}$)} & \colhead{($R_{\sun}$)} & \colhead{(cgs)} & \colhead{} & \colhead{($\mu$Hz)} & \colhead{($\mu$Hz)} &
\colhead{(K)} &\colhead{(cgs)} &\colhead{(cgs)}& \colhead{($M_{\sun}$)} & \colhead{($R_{\sun}$)}
}
\startdata
18 Del\tablenotemark{a}   & $  2.32\pm0.09$ & $ 8.32\pm0.14$ & $2.963\pm0.004$ & echelle  & 110.50 &   8.57& 5060 & $3.09\pm0.07$ & $ 2.82\pm0.06$ & $ 2.30$ & $ 8.50$ \\ 
24 Sex       & $  1.52\pm0.09$ & $ 5.21\pm0.11$ & $3.187\pm0.008$ & mean     & 185.83 &  14.02& 5020 & $3.40\pm0.08$ & $ 3.50\pm0.10$ & $ 1.54\pm0.08$ & $ 4.90\pm0.08$ \\ 
7 CMa        & $  1.49\pm0.07$ & $ 5.22\pm0.09$ & $3.175\pm0.007$ & mean     & 185.00 &  13.80& 4790 & $3.21\pm0.17$ & $ 3.19\pm0.06$ & $ 1.34\pm0.11$ & $ 4.87\pm0.17$ \\ 
$\alpha$ Ari&	--&	--&	--& --& --& -- & 4560 & $2.46\pm0.11$  & --  & --  & -- \\ 
BD+20 274    &    -- &   -- & $1.607\pm0.056$ & mean     &   5.25 &    --& 4360 & $1.75\pm0.10$ & $ 1.99\pm0.05$ & -- & -- \\ 
HD~100655     & $  1.50\pm0.08$ & $ 8.86\pm0.21$ & $2.720\pm0.006$ & echelle  &  64.50 &   6.28& 4850 & $2.78\pm0.08$ & $ 2.79\pm0.11$ & $ 2.28\pm0.69$ & $10.06\pm0.86$ \\ 
HD~102272     &    -- &  -- & $2.399\pm0.010$ & mean     &  31.00 &    --& 4790 & $2.54\pm0.08$ & $ 2.58\pm0.00$ & -- & -- \\ 
HD~102329&	--&	--&	-- & -- &--& --	 & 4780 & $3.01\pm0.09$  & --  &  -- & -- \\ 
HD~116029&	--&	--&	--& --& --& --	 & 4880 & $3.31\pm0.12$  & --  & --  & -- \\ 
HD~11977      &    -- &  -- & $2.678\pm0.008$ & mean     &  57.83 &    --& 4980 & $2.76\pm0.05$ & $ 2.90\pm0.20$ & $ 1.91\pm0.21$ & $10.09\pm0.32$ \\ 
HD~1690& 	--&	--&	--& --& --& --	 & 4370 & $1.91\pm0.13$  & --  & --  & -- \\ 
HD~18742      & $  2.23\pm0.08$ & $ 7.41\pm0.11$ & $3.047\pm0.004$ & mean     & 135.00 &  10.00& 4990 & $3.17\pm0.06$ & $ 3.09\pm0.02$ & $ 1.36\pm0.24$ & $ 5.13\pm0.23$ \\ 
HD~212771     & $  1.50\pm0.06$ & $ 4.66\pm0.07$ & $3.277\pm0.006$ & mean     & 227.00 &  16.44& 5100 & $3.52\pm0.09$ & $ 3.31\pm0.01$ & $ 1.56\pm0.18$ & $ 5.27\pm0.15$ \\ 
HD~28678      & $  1.48\pm0.07$ & $ 6.53\pm0.14$ & $2.980\pm0.004$ & mean & 115.00 &   9.87& 5050 & $3.15\pm0.07$ & $ 3.06\pm0.04$ & $ 1.53\pm0.51$ & $ 6.48\pm0.46$ \\ 
HD~30856      & $  1.16\pm0.07$ & $ 4.87\pm0.11$ & $3.128\pm0.007$ & mean     & 163.00 &  13.55& 4970 & $3.31\pm0.08$ & $ 3.20\pm0.08$ & $ 1.17\pm0.31$ & $ 4.40\pm0.17$ \\ 
HD~33142      & $  1.29\pm0.06$ & $ 4.08\pm0.07$ & $3.326\pm0.006$ & mean     & 256.50 &  18.59& 5000 & $3.50\pm0.11$ & $ 3.40\pm0.07$ & $ 1.41\pm0.31$ & $ 4.45\pm0.14$ \\ 
HD~4313      &	--&	--&	--&-- &-- & -- & 5000 &  $3.40\pm0.10$ & --  &  -- & -- \\ 
HD~5319       & $  1.33\pm0.07$ & $ 4.55\pm0.10$ & $3.247\pm0.006$ & mean     & 215.17 &  16.07& 4930 & $3.46\pm0.12$ & $ 3.26\pm0.06$ & $ 1.27\pm0.30$ & $ 4.06\pm0.42$ \\ 
HD~96063      & $  1.52\pm0.06$ & $ 4.40\pm0.06$ & $3.334\pm0.005$ & mean     & 259.50 &  18.07& 5060 & $3.29\pm0.11$ & $ 3.33\pm0.06$ & $ 1.37\pm0.07$ & $ 4.75\pm0.10$ \\ 
HD~98219      & $  1.29\pm0.05$ & $ 4.02\pm0.06$ & $3.342\pm0.005$ & echelle  & 268.00 &  19.08& 4930 & $3.36\pm0.09$ & $ 3.36\pm0.07$ & $ 1.41\pm0.31$ & $ 4.60\pm0.15$ \\ 
$\epsilon$ Tau     & -- & -- & $2.707\pm0.006$ & mean     &  62.50 &    --& 4860 & $2.71\pm0.09$ & $ 2.66\pm0.04$ & $ 2.57\pm0.21$ & $12.35\pm0.44$ \\ 
$\epsilon$ Ret   & $  1.31\pm0.05$ & $ 3.76\pm0.05$ & $3.405\pm0.005$ & mean   & 313.50 &  21.20& 4810 & $3.45\pm0.17$ & $ 3.78\pm0.06$ & $1.23$ & $3.18\pm0.08$ \\ 
\enddata
\tablenotetext{a}{Data became available after our original cutoff date but was included to allow for an expanded literature comparison in Section \ref{sec:validate}}
\end{deluxetable*}
\end{rotatetable*}

\begin{deluxetable*}{lrrrrrr|rr|}
\tablenum{4}
\tablecaption{Control Sample Final Results\label{tab:results2}}
\tablewidth{0pt}
\tablehead{
\colhead{} & \multicolumn{6}{c}{Asteroseismic} & \multicolumn{2}{c}{Spec}\\ 
\colhead{Star} & \colhead{Mass} & \colhead{Radius} & \colhead{$\log g$} & \colhead{Method} & \colhead{$\nu_{\rm max}$} & \colhead{$\Delta\nu$} &
\colhead{$T_{\rm eff}$} &\colhead{$\log g$} \\
\colhead{} & \colhead{($M_{\sun}$)} & \colhead{($R_{\sun}$)} & \colhead{(cgs)} & \colhead{} & \colhead{($\mu$Hz)} & \colhead{($\mu$Hz)} &
\colhead{(K)} &\colhead{(cgs)} 
}
\startdata
HD~105096& --& --& --& --& --& -- & 4750 &  $3.11\pm0.12$  \\ 
HD~108991     & $  1.56\pm0.08$ & $ 8.23\pm0.17$ & $2.800\pm0.005$ & mean     &  78.00 &   7.14& 4790 & $3.03\pm0.09$\\ 
HD~115202     & $  1.07\pm0.04$ & $ 5.18\pm0.07$ & $3.040\pm0.004$ & mean     & 134.95 &  11.86& 4830 & $3.20\pm0.10$\\ 
HD~121056     & $  1.20\pm0.05$ & $ 5.59\pm0.09$ & $3.023\pm0.005$ & mean     & 130.50 &  11.21& 4790 & $3.21\pm0.08$\\ 
HD~121156     & $  1.35\pm0.05$ & $ 6.54\pm0.11$ & $2.937\pm0.005$ & mean     & 108.50 &   9.39& 4670 & $3.07\pm0.15$\\ 
HD~17311&	--&	--&	--& --& -- & -- &5020 & $3.47\pm0.07$   \\ 
HD~19180&	--&	--&	--& --& -- & -- &4860 & $3.27\pm0.09$   \\ 
HD~205478     & $  1.45\pm0.15$ & $ 5.71\pm0.20$ & $3.087\pm0.014$ & mean     & 149.75 &  11.94& 4880 & $3.29\pm0.12$\\ 
HD~20924&	--&	--&	--& --& --& -- & 4680  & $3.00\pm0.14$   \\ 
HD~213066&	--&	--&	--& --& --& -- & 5830  & $3.85\pm0.08$   \\ 
HD~21340&	--&	--&	--& --& --& -- & 4970  & $3.16\pm0.08$   \\ 
HD~219553&	--&	--&	--& --& --& -- & 4870  & $3.25\pm0.12$   \\ 
HD~233860&	--&	--&	--& --& --& -- & 4750  & $2.75\pm0.06$   \\ 
HD~24148&	--&	--&	--& --& --& -- & 4970  & $3.22\pm0.09$   \\ 
HD~25069      & $  1.34\pm0.06$ & $ 4.81\pm0.08$ & $3.199\pm0.007$ & mean     & 193.50 &  14.79& 4900 & $3.33\pm0.12$\\ 
HD~30128&	--&	--&	--& --& --& -- & 4990 & $3.11\pm0.09$   \\ 
HD~45433&	--&	--&	--& --& --& -- & 4390 & $2.25\pm0.19$   \\ 
HIP~39079     & -- & -- & $2.097\pm0.019$ & mean     &  16.00 &  --& 4470 & $2.20\pm0.12$\\ 
HD~58540      & $  1.08\pm0.05$ & $ 5.88\pm0.10$ & $2.930\pm0.005$ & mean     & 106.17 &   9.82& 4720 & $3.08\pm0.10$\\ 
HD~6037       & $  1.92\pm0.12$ & $10.46\pm0.27$ & $2.681\pm0.006$ & mean     &  60.50 &   5.53& 4610 & $2.93\pm0.18$\\ 
HD~64152&	--&	--&	--& --& --& -- & 5070 & $3.14\pm0.07$  \\ 
HD~72292&	--&	--&	--& --& --& -- & 4490 & $2.39\pm0.19$   \\ 
HD~94386      & $  1.46\pm0.08$ & $ 8.59\pm0.19$ & $2.734\pm0.006$ & echelle  &  68.50 &   6.48& 4590 & $2.91\pm0.19$\\ 
HD~98579      & $  1.39\pm0.10$ & $ 8.89\pm0.24$ & $2.682\pm0.008$ & mean     &  60.50 &   6.00& 4630 & $2.78\pm0.14$\\ 
HIP~117756&	--&	--&	--& --& --& -- &  4620 &$2.51\pm0.10$  \\ 
Tyc0683-01190-1&	--&	--&	--& --& --& -- & 4710 &$2.53\pm0.09$   \\ 
\enddata
\end{deluxetable*}

\section{Results}  \label{sec:results}

Using the \texttt{Lightkurve} results from the TESS data, we were able to measure both $\Delta \nu$ and \numax\ for 23 of the stars in the sample:  13 of the planet hosts and 10 of the control.  For an additional 5 stars (4 planet hosts and 1 non-host), we measure only \numax. For 5 planet hosts 
and 15 control stars, we could not reliably infer either asteroseismic characteristics from the available light curves.
Using our measured \numax\ and  $\Delta\nu$ together with the spectroscopic $T_{\rm eff}$ \citep[from][]{2020AAS...23527403B},  we calculated the $\log g$, stellar mass and stellar radius  using the standard asteroseismic scaling relations \citep[][reproduced in Equations \ref{eq:1}-\ref{eq:3} below]{1991ApJ...368..599B,1995A&A...293...87K,1986ApJ...306L..37U}. 
\texttt{Lightkurve} adopts solar values of $\nu_{\rm max,\odot}=3090$~$\mu$Hz and $\Delta \nu_{\odot}=135.1$~$\mu$Hz from \cite{2011ApJ...743..143H}, while $T_{\rm eff,\odot}=5772$~K is adopted from \cite{2016AJ....152...41P}. The solar $\log g$ is derived from \texttt{astropy.constants}'s values of solar mass and radius, yielding $\log g_\odot=4.438$.
\begin{equation}
\label{eq:1}
    \frac{g}{g_\odot} = (\frac{\nu_{\rm max}}{\nu_{\rm max}\odot})(\frac{T_{\rm eff}}{T_{\rm eff}\odot})^{1/2}
\end{equation}
\begin{equation}
\label{eq:2}
    \frac{M}{M_\odot} = (\frac{\nu_{\rm max}}{\nu_{\rm max}\odot})^3(\frac{\Delta\nu}{\Delta\nu_\odot})^{-4}(\frac{T_{\rm eff}}{T_{\rm eff}\odot})^{3/2}
\end{equation}
\begin{equation}
\label{eq:3}
 \frac{R}{R_\odot} = 
(\frac{\nu_{\rm max}}{\nu_{\rm max}\odot})(\frac{\Delta\nu}{\Delta\nu_\odot})^{-2}(\frac{T_{\rm eff}}{T_{\rm eff}\odot})^{1/2}
\end{equation}

Our asteroseismic measurements along with the derived stellar characteristics are given in the first 7 columns of Tables \ref{tab:results} and \ref{tab:results2} for the planet host and control samples, respectively. 
For stars with more than one measurement, we used the echelle diagrams to discard any measurements without obvious vertical structure. If only one valid measurement remained, it was adopted as the solution and the method is identified as ``echelle''.  When more than one solution remained, we first averaged the available \numax\ and $\Delta \nu$ before calculating $\log g$, mass, and radius. These are identified 
in Tables \ref{tab:results}-\ref{tab:results2} with ``mean'' as the 
method.

\begin{figure}
\centering
\includegraphics[width=0.48\textwidth]{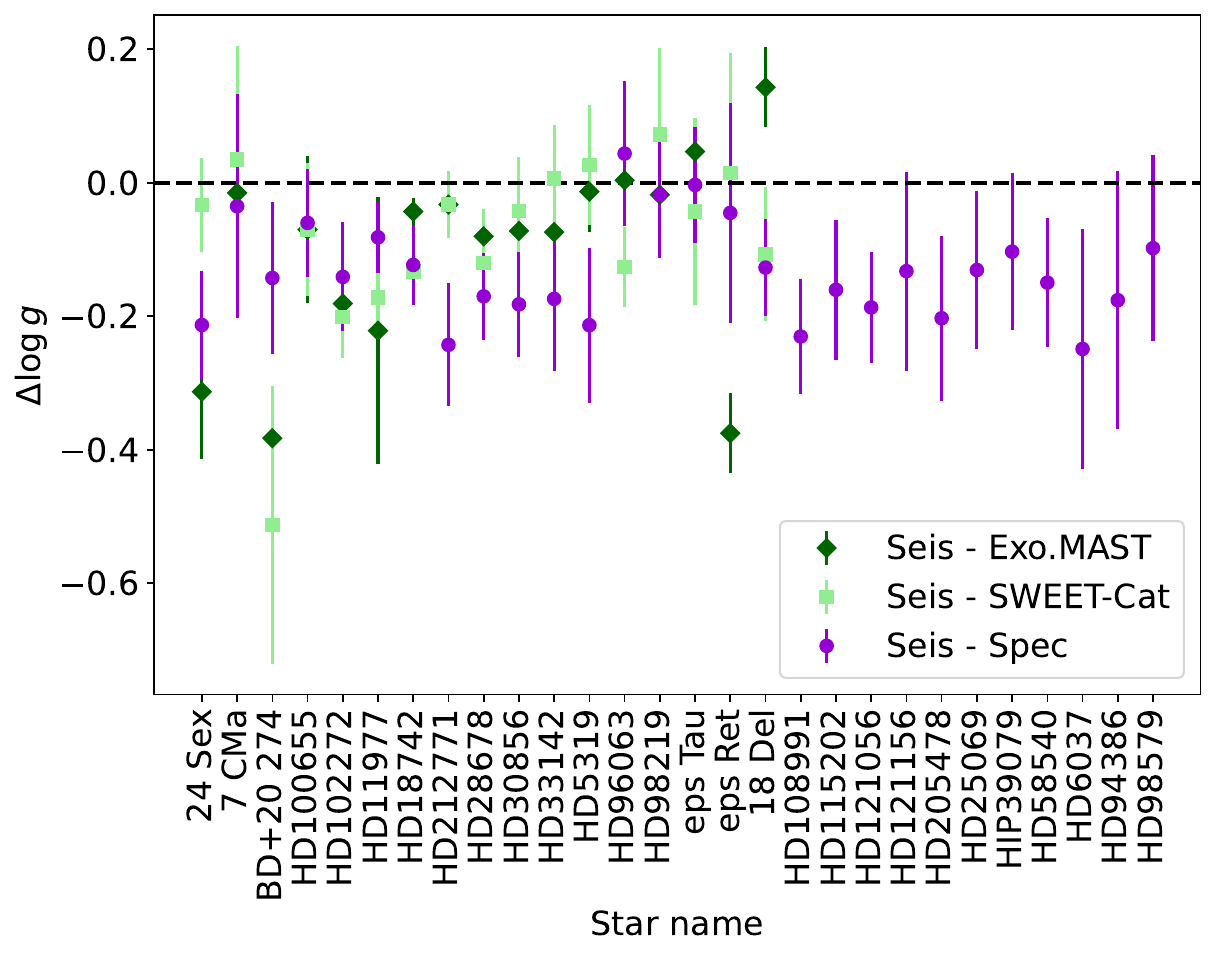}
\includegraphics[width=0.48\textwidth]{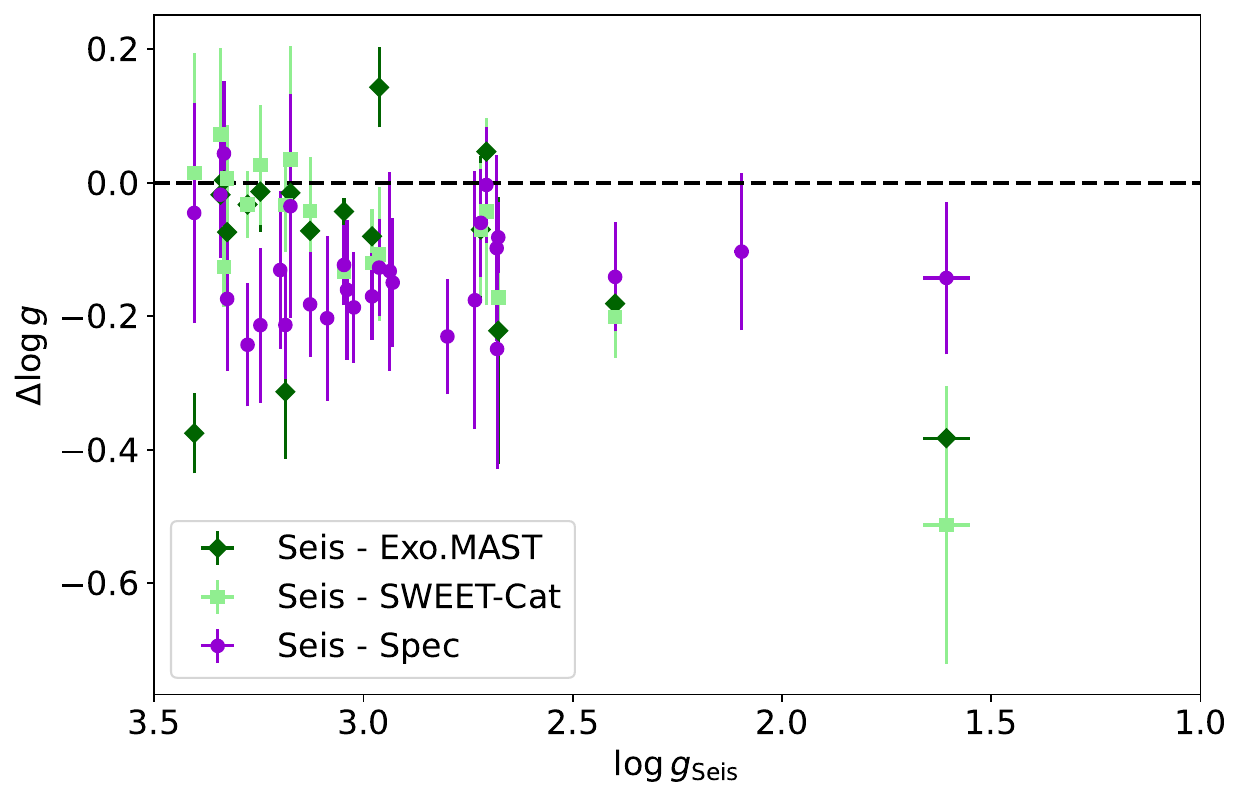}
\caption{Difference between the asteroseismic $\log g$ measured in this work and three different sources of spectroscopically measured $\log g$.
For the purple circles (labeled ``Seis-Spec''), the spectroscopic $\log g$ have been measured homogeneously by \cite{2020AAS...23527403B}.
Two additional literature sources available for the planet hosts are Exo.MAST (dark green diamonds) and SWEET-Cat (light green squares).
 The difference is plotted both against stellar name (top) and asteroseismic $\log g$ (bottom).}
\label{fig:logg}
\end{figure}

Next, we compare our asteroseismically derived $\log g$ to literature measurements, which is shown in Figure \ref{fig:logg}.  All stars in this work have $\log g$ measured by \citet[][identified as ``Spec'' in the Figure]{2020AAS...23527403B}, and the planet hosts have additional $\log g$ available from both Exo.MAST (with different primary sources) and SWEET-Cat. 
There is a tendency for the asteroseismic $
\log g$ to be consistently lower than the spectroscopic ones (by 0.1~dex on average),  a difference that exceeds the formal uncertainties in the majority of cases. Figure \ref{fig:logg} shows this difference as a function of the asteroseismic $\log g$. For the Exo.MAST and ``Spec'' samples, there is no indication that the discrepancy depends on the stellar $\log g$.
However, there is a dependence when comparing with SWEET-Cat.

\begin{figure*}
\includegraphics[width=0.48\textwidth]{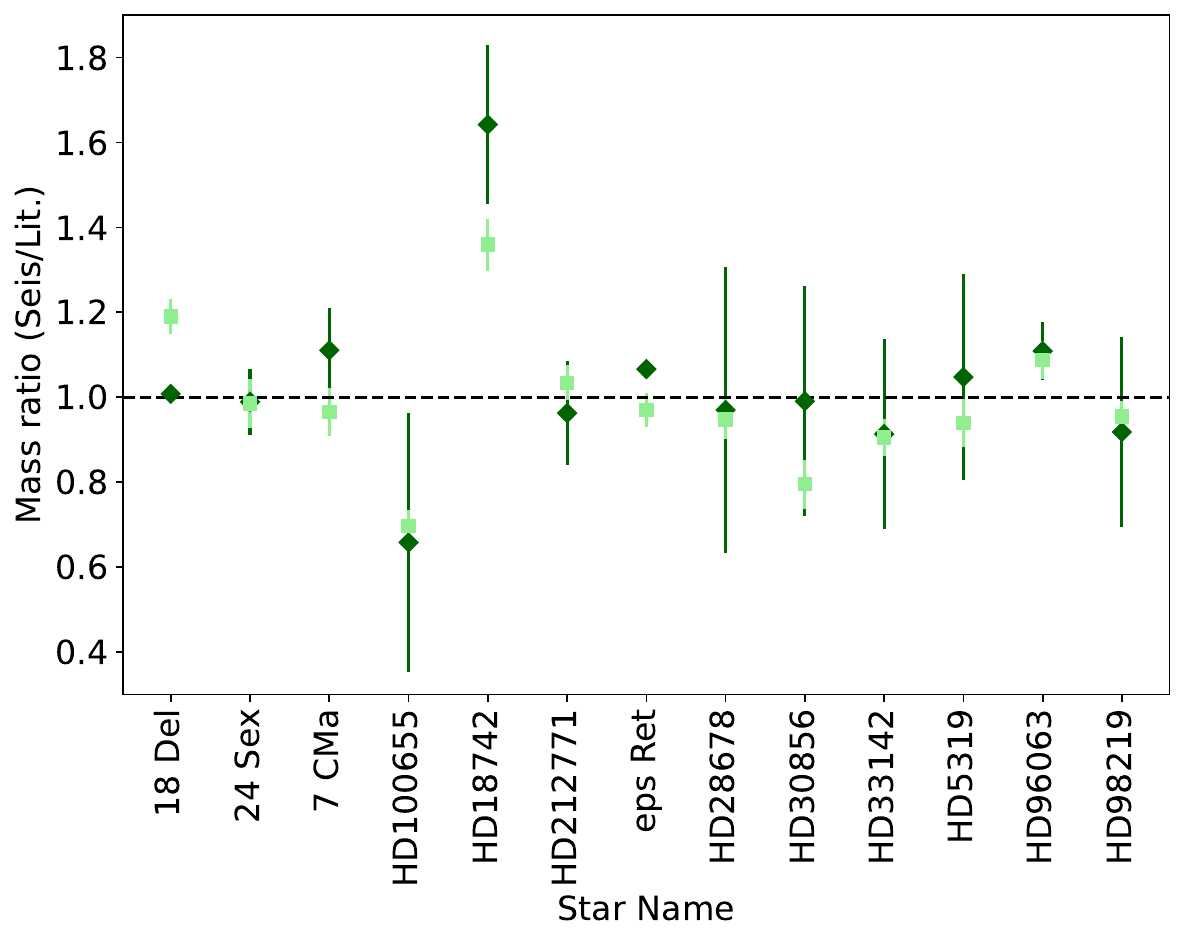}
\includegraphics[width=0.48\textwidth]{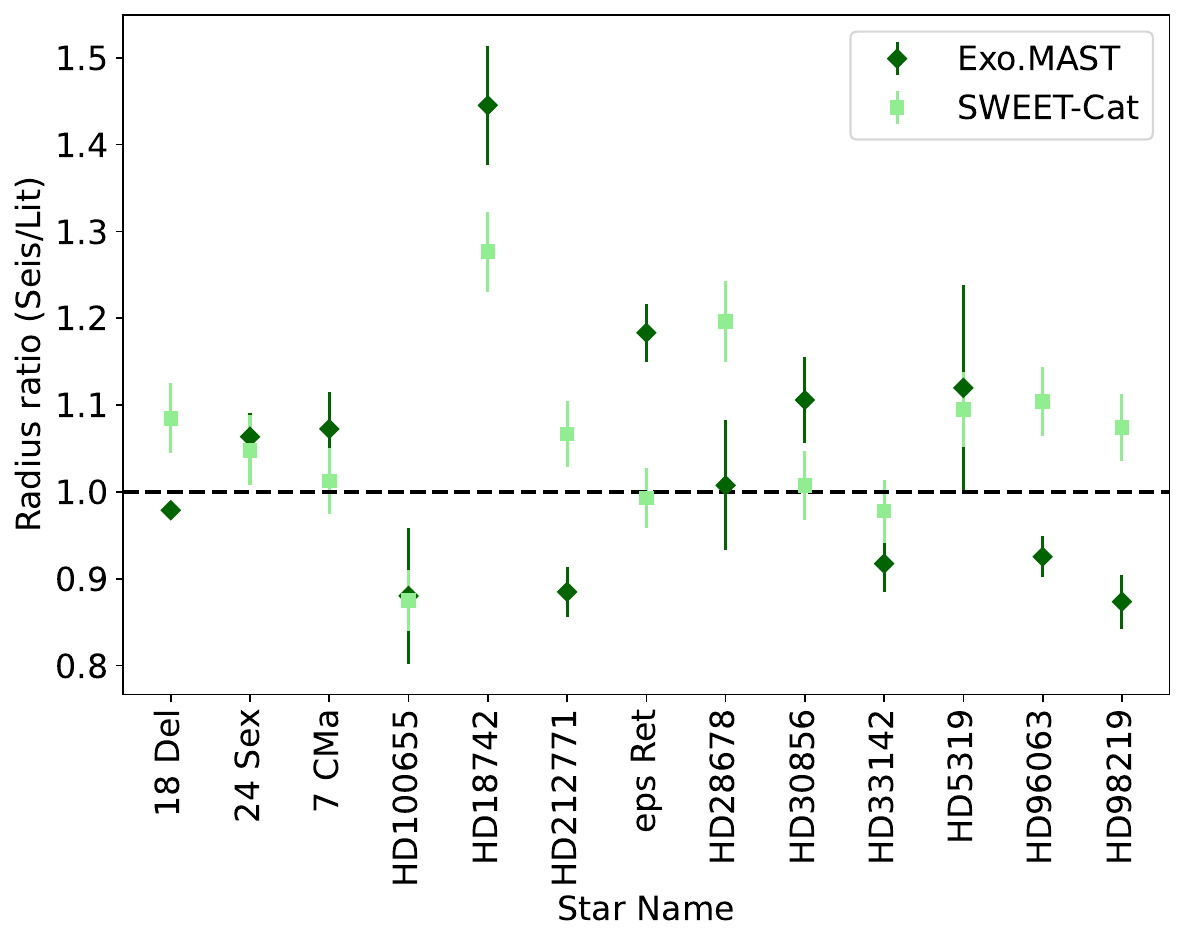} 
\includegraphics[width=0.48\textwidth]{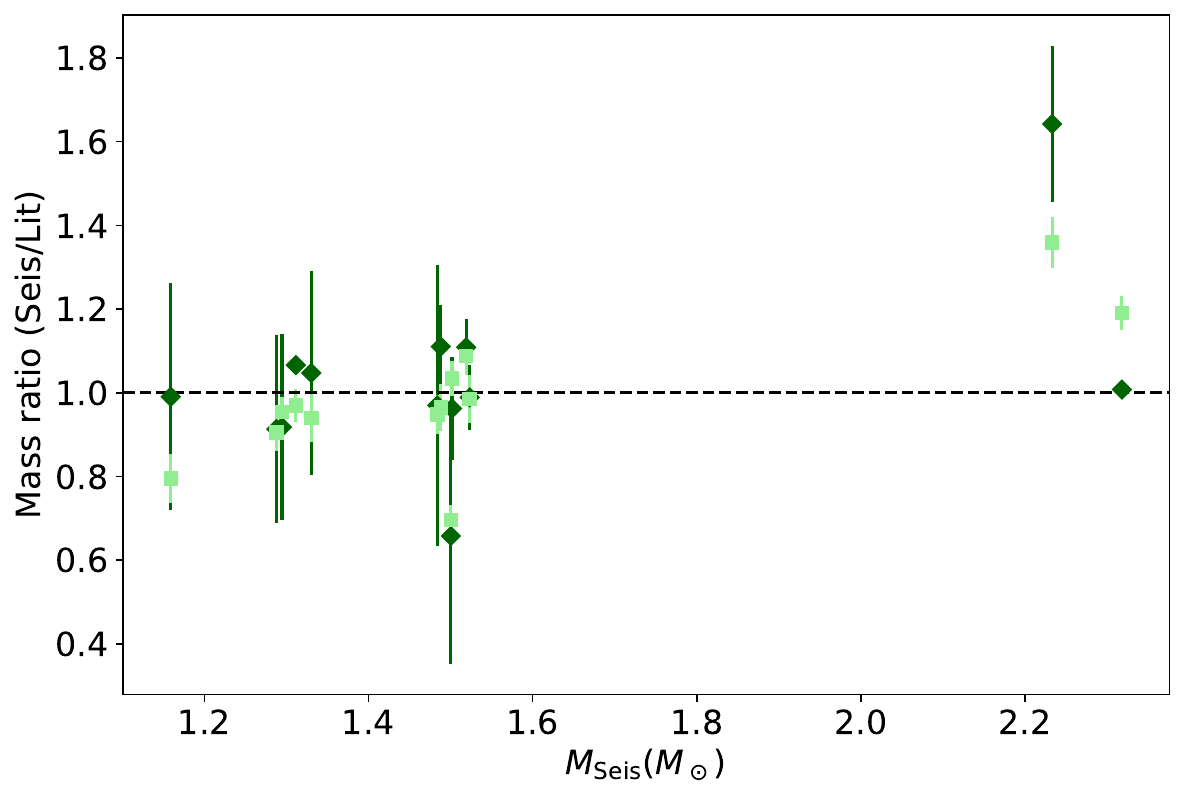}
\includegraphics[width=0.48\textwidth]{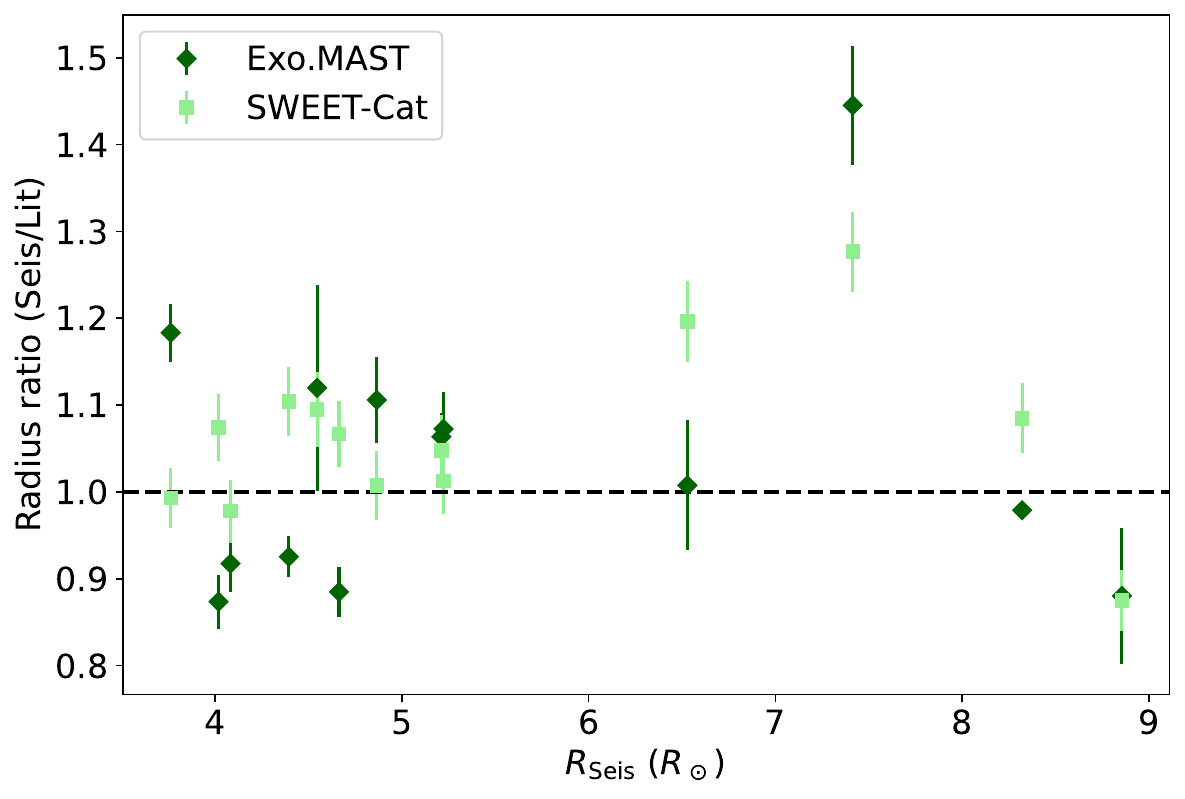}
\caption{Ratio of our asteroseismically measured masses (left) and radii (right) compared to those reported in Exo.MAST (dark diamonds) and in SWEET-Cat (light squares).} Comparisons are again plotted both to identify individual stars (top) and against our asteroseismic measurements (bottom). Note that no uncertainties are reported for 18 Del or for $\epsilon$~Ret's mass and that for HD33142, the mass and the radius came from different sources in Exo.MAST.\label{fig:mass}\end{figure*}

A similar comparison was done on the masses and radii, as shown in Figure \ref{fig:mass}. This comparison is restricted to the planet host stars, for which previously measured masses and radii are readily available.
The systematic offset that was present in the $\log g$ did not translate to the masses.  Most of the asteroseismic masses were relatively close to the Exo.MAST masses and within $1\sigma$ of the uncertainty, with no clear preference for the asteroseismic masses to be larger or smaller than those in the literature.  
The SWEET-Cat masses showed similar size discrepancies with the asteroseismic masses but with significantly smaller reported error bars than Exo.MAST. The discrepancies between the asteroseismic and SWEET-Cat masses also show a positive trend with the seismic mass. Stars more massive than $\sim$1.5~\msun\ have larger asteroseismic masses relative to SWEET-Cat (by up to 40\%), while stars less massive than $\sim$1.5~\msun\ have smaller asteroseismic masses (by up to 20\%).
The stellar radii comparison, on the other hand, show an unanticipated result. Almost none of the asteroseismic radii agreed with the Exo.MAST radii within the quoted uncertainties, with half being  higher by $\sim 10$\%, and half $\sim 10$\% lower. Those with  larger asteroseismic radii are qualitatively consistent with the typically lower asteroseismic $\log g$ seen in Figure \ref{fig:logg}, since the radius measurement is inversely proportional to the $\log g$. However, a 10\% change in radius corresponds to a $\sim$0.08~dex change in $\log g$, whereas Figure \ref{fig:logg} shows $\log g$ differences of either $\sim$0.05~dex or $\geq$0.15~dex for the Exo.MAST $\log g$'s (dark green points). This apparent paradox can be understood if one calculates $\log g$ directly from the reported Exo.MAST masses and radii and compare them to the reported $\log g$. We find a mean absolute difference between these two $\log g$'s to be 0.10~dex, with individual discrepancies as large as 0.25~dex. Compare this to the typical reported $\log g$ uncertainty of 0.06~dex and maximum uncertainty of $0.20$~dex.
The comparison between the asteroseismic radii and SWEET-Cat radii showed overall better agreement, though there was a more systematic tendency for the asteroseismic radii to be larger.

\subsection{Asteroseismic Systematics}
The analysis in this work relies on the simple application of scaling relations tied to the Sun that are known to accrue systematic inaccuracies as the stellar properties deviate from the solar values.  A number of papers have explored these deviations for red giant stars. For example, \cite{Li2022}  undertook an extensive study of \textit{Kepler} red giants with spectra available in the APOGEE and LAMOST surveys and presented revised scaling relations tailored to those surveys' pipeline determinations of $T_{\rm eff}$ and [M/H]. However, we cannot apply these corrected relations uniformly to our data because we have stars with parameters outside the fitted range. Additionally, to apply the corrections we also need to understand in detail any temperature offsets between our $T_{\rm eff}$ and those of APOGEE or LAMOST. Instead, we explore the impact of excluding these higher order corrections on our  results by adopting the results of \cite{2016MNRAS.460.4277G}, who parameterized the corrections as $T_{\rm eff}$ and [M/H] dependent adjustments to the reference $\Delta \nu$. We calculated adjustments for the stars in Tables \ref{tab:results} and \ref{tab:results2} for which we had a $\Delta \nu$ measurement. On average, the adjusted $\Delta \nu$ applicable to our results is 134.9 $\mu$Hz (range 133.1 --
136.7), compared to our adopted solar reference value of 135.1~$\mu$Hz. 
Stellar temperature is the dominant factor, and stars with $T_{\rm eff} \gtrsim 4800$~K (about $\sim 65$\% of the full sample) have overestimated masses and radii, while the stars with $T_{\rm eff} \lesssim 4800$~K  have underestimated masses and radii.
The median correction factor of the full sample for both mass and radii is 1.00, and the range of correction factors is 0.94--1.05 in mass and 0.97--1.03 in radius.  
The exoplanet sub-sample, on the other hand, predominantly contains hotter stars with overestimated parameters. The median correction factors for mass and radii are 0.97 and 0.98, respectively. 
 This up to 3\% correction factor in radius could account for some of the radius discrepancy between the asteroseismic and literature radii seen in  Figure \ref{fig:mass}.  However, for 5 stars (18~Del, HD~212771, $\epsilon$~Ret, HD~96063, and HD~98219), applying a correction would improve the agreement with one literature source while worsening it for the other.   Of the remaining stars, only 7 CMa has a correction that would increase our reported radius, but our reported value is already larger than both literature radii measurements. Similarly, for both HD~100655 and HD~33142, applying a correction will reduce the asteroseismic radius and worsen the discrepancy with both literature sources.  For the remaining five stars (24 Sex, HD~18742, HD~28678, HD~30856, and HD~5319), the correction will reduce, but generally not eliminate the 1\%--44\% discrepancies.

\subsection{Validation of Derived Parameters}
\label{sec:validate}
Several planet hosting stars have previously derived asteroseismic properties from independent analyses that we can compare our results to. \cite{2019ApJ...885...31C} analyzed early TESS photometry of HD~212771 (Sector 2) and HD~203949 (Sector 1), the former of which was re-analyzed in this work with additional TESS data. Table \ref{tab:compare} lists their asteroseismic results compared to our own. 
Our radii and theirs differ by only $\sim$1.5\%, well within the 4\% uncertainties. Both of these radii measurements are
$\sim 88$\% of the Exo.MAST value. 
The \cite{2019ApJ...885...31C} mass measurement is smaller than both our mass measurement (by 5\%) and the Exo.MAST mass measurement (by 9\%), though both are still consistent within their respective uncertainties. 

Additionally,  \cite{2017MNRAS.472.4110S} used time series radial velocity data at high precision ($< 3$ m s$^{-1}$) using the 1-m SONG telescope to measure the \numax\ for 8 red giants, including 18 Del and $\epsilon~Tau$. \cite{2017MNRAS.472.4110S} only measured \numax\ for these stars, and we find excellent agreement with their results.
\begin{deluxetable}{Lcc} %
\tablenum{5}
\tablecaption{Literature Comparison\label{tab:compare}}
\tablewidth{0pt}
\tablehead{
\colhead{Parameter} & \colhead{Literature} & \colhead{This Work} }
\startdata
\multicolumn{3}{c}{HD 212771\tablenotemark{a}} \\
\hline
\nu_{\rm max}\ (\mu {\rm Hz}) & $226.6\pm9.4$ & $227.6\pm2.5$\\
\Delta \nu \ (\mu {\rm Hz}) & $16.25\pm0.19$ & $16.45\pm0.06$\\
\log g & $3.263\pm0.01$ & $3.277\pm0.006$ \\
M_{\star}  (M_{\odot})&$1.42\pm0.07$ & $1.50\pm0.06$ \\
R_{\star} (R_{\odot}) & $4.61\pm0.09$ & $4.66\pm0.07$ \\
\hline
\multicolumn{3}{c}{18 Del\tablenotemark{b} } \\
\hline
\nu_{\rm max}\ (\mu {\rm Hz}) & $112\pm17$ & $110.5\pm0.7$ \\
\log g &$ 2.97 \pm0.09$& $2.963\pm0.004$ \\
\hline
\multicolumn{3}{c}{$\epsilon$~Tau\tablenotemark{b} } \\
\hline
\nu_{\rm max}\ (\mu {\rm Hz}) & $56.9\pm8.5$ & $62.5\pm0.7$ \\
\log g &$ 2.67\pm0.08$ & $2.707\pm0.006$ \\
\enddata
\tablenotetext{a}{\cite{2019ApJ...885...31C}}
\tablenotetext{b}{\cite{2017MNRAS.472.4110S}}
\end{deluxetable}

Finally, we compared our list of giants to the \cite{hon_quick_2021} study of 158,000 TESS red giants processed through an automated pipeline that only measured \numax, and we found  9 of our targets in that study. \cite{hon_quick_2021} used Gaia parallaxes and photometry in combination with \numax\ to constrain effective temperature and radii. A comparison of these results are given in Table \ref{tab:compare2}. Large discrepancies exist with the derived temperatures, with the spectroscopic temperatures being larger in all cases. However, although \cite{hon_quick_2021} found statistically that their temperatures were consistent with spectroscopically-measured  temperatures from APOGEE, the largest disagreements (more than $\sim$100~K) occurred for stars with high spectroscopic temperatures but low photometric temperatures. Despite the temperature disagreement, we find that all of our radii measurements agree within the uncertainties reported by  \cite{hon_quick_2021} with the exception of HD~100655, which is discussed in more detail in Section \ref{sec:outlier} below. 

\begin{deluxetable*}{Lr|ccc|ccc} %
\tablenum{6}
\tablecaption{\cite{hon_quick_2021} Comparison\label{tab:compare2}}
\tablewidth{0pt}
\tablehead{
\colhead{}& \colhead{} & \multicolumn{3}{c}{Literature} & \multicolumn{3}{c}{This Work}\\ 
\colhead{Name}& \colhead{TIC ID} &
\colhead{$T_{\rm eff}$} & \colhead{\numax} & \colhead{$R_\star$} &
\colhead{$T_{\rm eff}$} & \colhead{\numax} & \colhead{$R_\star$} \\
\colhead{} &\colhead{} &\colhead{(K)} &\colhead{($\mu$Hz)} &\colhead{($R_\odot$)} &
\colhead{(K)} &\colhead{($\mu$Hz)} &\colhead{($R_\odot$)} }
\startdata
{\rm HD~108991}&	130867823	& $4264\pm85$	& $75.6	\pm4.04$ & $7.9	\pm0.4$     &4790 &	78.00	& 8.23 \\
{\rm HD~115202}&	422432907	& $4637\pm92$	& $140.2\pm4.74$ & $5.3	\pm0.6$     &4830 &	134.95	& 5.18 \\
{\rm HD~121056}&	111947706	& $4515\pm90$	& $129.7\pm5.94$ & $6.0	\pm0.7$     &4790 &	130.50	& 5.59 \\
{\rm HD~121156}&	72593986	& $4486\pm89$	& $113.6\pm5.74$ & $6.3	\pm0.9$     & 4670&	108.50	& 6.54 \\
{\rm HD~25069}	&    49952922	& $4731\pm94$	& $197.1\pm5.94$ & $4.6	\pm0.6$ &4900 &	193.50	& 4.81 \\
{\rm HD~98579}	&    375563752	& $4489\pm89$	& $62.1	\pm3.64$ & $8.7	\pm1.2$ &4630 &	60.50	& 8.89 \\
{\rm HD~100655}&	156942082	& $4064\pm81$	& $60.7	\pm3.74$ & $11.6	\pm0.5$ &4850 &	64.5	& 8.86 \\
{\rm HD~102272}&	82605074	& $4754\pm95$	& $26.8	\pm2.24$ & $10.1	\pm0.4$ & 4790 &	31.0	& -- \\
{\rm HD~28678}	&   449145115	& $5021\pm100$&$126.6	\pm9.6$ & $6.7	\pm0.3$ & 5050&	115.0	& 6.53 \\
\enddata
\end{deluxetable*}

\subsection{Outliers}
\label{sec:outlier}
Two of the planet hosts (HD 100655 and HD 18742) are significant outliers in both the mass and radius comparison plots, but do not particularly stand out as outliers in the $\log g$ comparison. We revisited the asteroseismic analysis of these stars to see whether these were near our detection threshold. HD~18742 presented no such concerns, and we are confident in our analysis. For HD~100655, the $\Delta \nu$ measurement required using \texttt{Lightkurve}'s echelle diagram diagnostic to discriminate between two different possibilities, but the masses derived from either $\Delta \nu$ measurement were both significantly smaller than the non-asteroseismically derived masses. The rejected $\Delta \nu$ solution of 6.59~$\mu$Hz corresponds to an even smaller stellar mass of 1.23~$M_\odot$.

\begin{figure}
    \centering
    \includegraphics[width=0.5\textwidth]{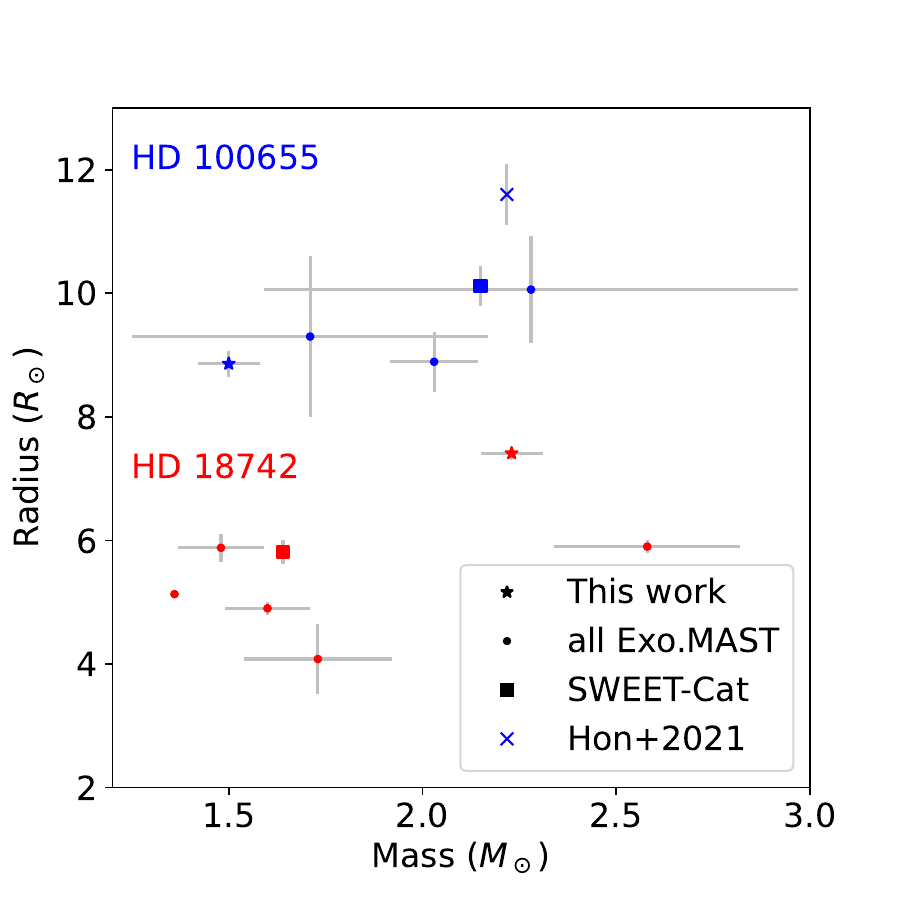}
    \caption{Comparison of our asteroseismic masses and radii (stars) compared to various non-asteroseismic literature sources (circles), SWEET-Cat (squares), and joint asteroseismic (\numax\ only) and Gaia-based measurement (x) from \cite{hon_quick_2021}.
    \label{fig:outlier}}
\end{figure}

For these two stars, we looked at the full array of measured masses and radii available at the NASA Exoplanet archive, instead of just the default values returned by Exo.MAST. The comparisons are shown in Figure \ref{fig:outlier}, where the star-symbols are our asteroseismically measured values and the circles are from the literature. The SWEET-Cat values are additionally plotted as squares.
For HD~100655, the upper right circle \citep[from][]{2017AJ....153..136S} was the default value returned by Exo.MAST. Additional data come from \cite{2012PASJ...64...34O} and \cite{2015A&A...576A..94S}. For HD~18472, the default is the left most (no error bar) point \citep[from][]{2019AJ....157..149L}, and additional data points are from \cite{2011ApJS..197...26J,2015A&A...574A..50J,2017AJ....153..136S} and \cite{2013A&A...557A..70M}. For both stars, there are alternative literature measurements in better agreement with our asteroseismic results. For HD~100655 we additionally plot the asteroseismic/Gaia measurement from \cite{hon_quick_2021} using their Equation~1 to derive the mass. We note that \cite{hon_quick_2021} only used the derived mass as a consistency check, and HD~100655 had the largest temperature discrepancy in Table \ref{tab:compare2}. The temperatures obtained from Exo.MAST are $4801\pm60$ and $4891\pm46$ and are consistent with our own spectroscopically derived temperature.
HD~18472 is notable for the two apparent families of solutions. However, even within the lower mass family of solutions, the results diverge outside the formal uncertainties. We conclude, therefore, that these outliers are indicative of the lingering difficulty of constraining these properties for some red giant stars.

\section{Potential Impacts on Planet Studies}
\subsection{Revisiting the ``Retired A Stars''}
\begin{figure}
    \centering
    \includegraphics[width=0.5\textwidth]{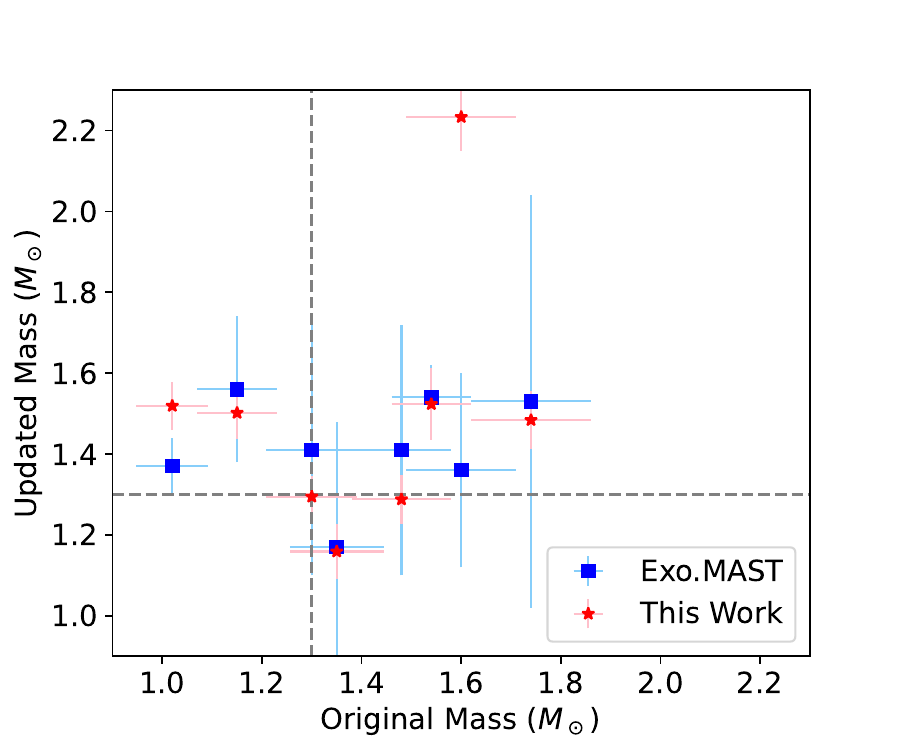}
    \caption{Comparison of the originally published masses of intermediate mass stars from the ``Retired A Stars'' survey in our sample, compared to the updated values derived in this work with asteroseismology (red stars) or in the literature (as accessed through Exo.MAST, blue squares). Dotted lines at 1.3~\msun\ separate intermediate masses from low masses.}
    \label{fig:retiredA}
\end{figure}

Eight of the planet hosts for which we have new asteroseismic masses in Table \ref{tab:results} are from the  ``Retired A Star'' sample and had stellar characteristics originally published in \cite{johnson_retired_2010_IV,johnson_retired_2011_VI,johnson_retired_2011_VII}.
They are HD~212771, 24~Sex, HD~18742, HD~28678, HD~30856, HD~33142, HD~96063, and HD~98219.
The target mass demographic of this survey was $M_\star > 1.3$~\msun\ \citep{johnson_eccentric_2006}, and (as discussed in Section \ref{sec:intro}) has been the subject of controversy in the literature. Figure \ref{fig:retiredA} compares the originally published stellar masses to both our asteroseismically measured masses and the literature masses we retrieved from Exo.MAST. We find that the most massive stars generally have slightly lower updated masses compared to the original derivations. Nevertheless, all but one (HD~30856) has a revised mass  within the targeted intermediate mass range. Furthermore, the two planet hosts with originally published masses \textit{below} 1.3~\msun\ have revised masses that place them in the intermediate mass regime.  The  extreme outlier  is HD~18742, previously discussed in Section \ref{sec:outlier}.

\subsection{Impact to Inferred Planetary Properties}

Planetary properties and their uncertainties are inextricably tied to the host star parameters, and here we explore impacts of revised stellar properties. In transiting systems, the observable (transit depth) scales with $R_{\rm p}^2/R_{\star}^2$, so there is a linear dependence on the derived planet radii relative to the adopted stellar radii.
By measuring the stellar Doppler motion induced by an orbiting planet, the minimum planetary mass can be inferred from $M_{\rm p}\sin(i) \sim M_\star^{2/3}K_\star P^{1/3} $, where $P$ is the observed orbital period and $K_\star$ is the radial velocity semi-amplitude.

Our comparison of literature masses to those derived from asteroseismology in Figure \ref{fig:mass} suggests relatively good agreement with those available in Exo.MAST, and  importantly, that quoted uncertainties are representative of those deviations. The SWEET-Cat masses, in contrast, may have slightly underestimated uncertainties and a slight systematic in derived masses.These points are true of planetary masses that depend on them. Some caution is warranted given that some individual stars have had substantial revisions or outstanding discrepancies in their masses  (e.g., Figures \ref{fig:outlier}  and \ref{fig:retiredA}). However, the impacts of uncertain stellar masses to planetary masses is mitigated by the fact that for our sample of RV detected systems the planetary masses  are minimums only, due to unknown inclinations.

Our results are therefore most likely impactful in the interpretation of planets in transiting systems (where the true planetary mass is measured and where planetary radii can be inferred) for stars that are similarly characterized with non-asteroseismic methods. Even for exoplanets discovered around evolved stars via transits using TESS data \citep[e.g.,][]{GTGI,GTGII,GTGIII}, it is still common to utilize some variation of isochrone-fitting to model grids to constrain the stellar mass and radius, especially if the available TESS data is insufficient for direct asteroseismic analysis (as was true for $\sim$50\% of our sample data.) 
Errors in the stellar radius would lead to correspondingly large errors in the planetary radius. 

We found that the discrepancies between the asteroseismic and non-asteroseismic determinations of stellar radii differ
by $\sim 10$\%.  In the Exo.MAST comparison, only two stars (assuming a typical uncertainty for 18 Del) have radii that agree within the uncertainties, and the scatter goes in both directions independent of the size of the star.
The SWEET-Cat catalog showed a tendency to  underestimate radii relative to asteroseismology, with most of 
these discrepancies exceeding the reported uncertainties, again underscoring an overconfidence in reported parameters. The overconfidence in  uncertainties is, perhaps, unsurprising. \cite{Tayar2022} recently estimated a minimum floor of $\sim4$\% uncertainty in radius in most optimistic scenarios (when angular diameter measurements are available). Compare this to the median quoted uncertainty  $\sim3$\% in the literature for our sample. 

\begin{figure}
    \centering
    \includegraphics[width=0.45\textwidth]{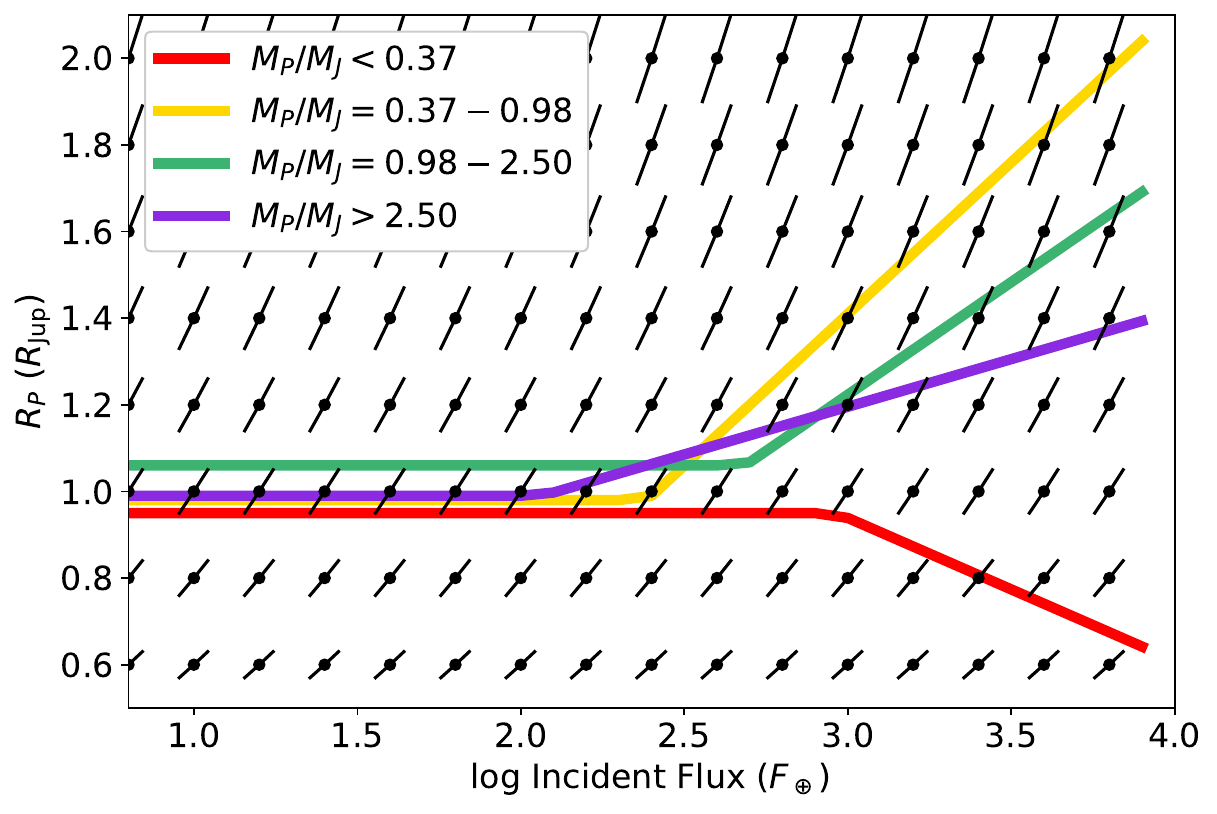}
    \caption{Solid, colored lines show the empirical trend of exoplanet radii as a function of the incident stellar flux on the planet in different planet mass bins from \cite{2018A&A...616A..76S}. The black points with error bars shows how a $\pm 5\%$ error in the stellar radius affects the determination of the planetary radius and incident flux.}
    \label{fig:planet_flux}
\end{figure}

As an example, we explore the implications of revised  stellar radii on the degree of planetary inflation due to irradiation \citep{2016ApJ...818....4L}. The stellar radius is used in the determination of both the planetary radius and the incident flux on the planet (F). The latter goes as $(R_\star/a)^2\sigma T^4_\star$, where $T_\star$
is the stellar temperature, $a$ is the orbital separation, and $\sigma$ is the Stefan-Boltzmann constant. \cite{2018A&A...616A..76S} fit empirical relationships  between stellar irradiation and planet radii  in different mass bins, and they
 found that above some threshold flux, the more massive planets' radii  increase linearly with the logarithm of the incident flux. In the smallest mass bin, the radii decrease with increasing flux. The solid lines in Figure \ref{fig:planet_flux} show these relationships. The grid of points with error bars shows the impact that a $\pm 5\%$ error in the adopted stellar radius would have on the data underlying these fits. The planetary radii will shift proportionally in the vertical direction, but the calculated incident flux will shift by a near constant value of $\pm 0.04$~dex in this example. The impact to understanding inflation will therefore depend both on where the planet falls in this parameter space and on the underlying physical relationship. In general, it increases the dispersion around the underlying relationship, but the effect is mitigated in regions where the shift is parallel to the true underlying trend, and more accentuated where it is perpendicular to it. This error could be misinterpreted as true dispersion and hamper the physical interpretation of the relationships.

Accurate stellar radii are also crucial for understanding the eventual demise of hot Jupiters, since the tidal decay of a planet's orbit is a steep function of $a/R_\star$ \citep{1995A&A...296..709V}. With the orders of magnitude expansion of the stellar radius during the red giant phase, it is well known that planets on the shortest period orbits will not survive the post main sequence  (or in some cases even the main sequence) lifetime of their host star before tidal decay leads to their engulfment \citep{2009ApJ...700..832C,2009ApJ...705L..81V,2011ApJ...737...66K}.  This engulfment could naturally explain the relative dearth of close-in planets found around evolved stars by radial velocity surveys, particularly those of red giants that include red clump stars \citep{2008PASJ...60..539S}. However, it would require stronger tidal interactions than expected to account for the dearth of planets around subgiants and low luminosity red giants that have not yet substantially increased swelled in size \citep{2010ApJ...709..396B}. More recent discoveries of planets around evolved stars via transits have begun filling in this parameter space \citep[beginning with \textit{Kepler} discoveries, e.g., with][]{lillobox2014,2015ApJ...800...46B} and continuing with TESS \citep{2019AJ....157..245H}. Both \cite{Grunblatt19}  and \cite{2023A&A...670A..26T} found the hot Jupiter frequency around evolved stars on the lower red giant branch derived from transiting surveys are consistent with the occurrence around main sequence stars.

It is of note that the \cite{Grunblatt19} study of hot Jupiter occurrence rate did utilize asteroseismology from the NASA \textit{K2} mission \citep{2014PASP..126..398H} to characterize the masses and radii of their $\sim 2500$ lower red giant branch sample. Their comparison of asteroseismic radii to two parallax-based methods did not yield a clear systematic offset between stellar radii. Nevertheless, of the three planets in their sample, their analysis resulted in a large increase in the radius of one the planets (K2-161) relative to its discovery radius by a factor of $\sim 3$.

\section{Conclusions}  
\label{sec:done}
We analyzed TESS light curves using the public \texttt{Lighkurve} package to measure asteroseismic properties of red giants that had been targeted by radial velocity surveys for finding planetary companions. 
In our red giant sample, we found that the spectroscopically measured $\log g$ tend to be systematically larger than those derived from asteroseismology. Such systematics have been found in previous comparisons between asteroseismic and spectroscopic 
$\log g$. While we expected this systematic to be reflected as systematically overestimated masses for red giant planet hosts,
we instead found that the asteroseismic masses were largely consistent with the literature. However, the stellar radii generally disagreed with literature measurements outside of the formal uncertainties. 

Our investigation of outliers suggests that there are still red giant parameter spaces where precise stellar parameters remain an elusive goal despite the wealth of readily available data.  Particular caution is advised when combining data from heterogeneous sources as formal uncertainties continue to underestimate systematics between different methodologies. We also 
caution that individual objects within large automated homogeneous analyses may require extra scrutiny, especially if they have unexpected properties, as large deviations can exist  in some situations even when the sample as a whole has statistically validated results.  The case of HD~100065 is a telling example, with a 800~K discrepancy between the automated \texttt{isoclassify} \citep{2017ApJ...844..102H,2020AJ....159..280B} temperature derived by \cite{hon_quick_2021} and multiple, independent spectroscopic temperature measurements.

\begin{acknowledgments}
JKC would like to thank Jamie Tayar for helpful discussions on identifying frequencies in the oscillation spectra and Mitch Revalski for help assembling software citations. The authors thank the anonymous referee for their thorough reviews and suggestions that improved the presentation of the work.  This research has made use of the NASA Exoplanet Archive, which is operated by the California Institute of Technology, under contract with the National Aeronautics and Space Administration under the Exoplanet Exploration Program.
This paper includes data collected with the TESS mission, obtained from the MAST data archive at the Space Telescope Science Institute (STScI). Funding for the TESS mission is provided by the NASA Explorer Program. This work was supported by the TESS Guest Investigator Program grant 80NSSC21K1025. STScI is operated by the Association of Universities for Research in Astronomy, Inc., under NASA contract NAS 5–26555.
\end{acknowledgments}

%

\vspace{5mm}
\facility{TESS}


\software{Astropy \citep{astropy:2013, astropy:2018, astropy:2022},  
          Astroquery \cite{ginsburg2019},
          Jupyter \citep{Kluyver2016}, 
          Lightkurve \citep{2018ascl.soft12013L}, 
          Matplotlib \citep{Hunter2007, Caswell2021}, 
          NumPy \citep{Harris2020}, 
          Pandas \citep{reback2020pandas,mckinney-proc-scipy-2010}, 
          Python (\citealp{VanRossum2009}, \url{https://www.python.org}), 
          Scipy \citep{Virtanen2020a, Virtanen2020b},
          Uncertainties \citep{lebigot}}




\bibliography{references,software_references}{}
\bibliographystyle{aasjournal}



\end{document}